\documentclass[acmtog]{acmart}
\acmSubmissionID{534}

\usepackage{multirow} 
\usepackage{subcaption}
\usepackage[ruled]{algorithm2e} 

\SetAlFnt{\small}
\SetAlCapFnt{\small}
\SetAlCapNameFnt{\small}
\SetAlCapHSkip{0pt}

\copyrightyear{2023} 
\acmYear{2023} 
\setcopyright{acmlicensed}
\acmConference[SA Conference Papers '23]{SIGGRAPH Asia 2023 Conference Papers}{December 12--15, 2023}{Sydney, NSW, Australia}
\acmBooktitle{SIGGRAPH Asia 2023 Conference Papers (SA Conference Papers '23), December 12--15, 2023, Sydney, NSW, Australia}
\acmPrice{15.00}
\acmDOI{10.1145/3610548.3618191}
\acmISBN{979-8-4007-0315-7/23/12}

\begin{document}
\title{A Hessian-Based Field Deformer for Real-Time Topology-Aware Shape Editing 
}



\author{Yunxiao Zhang}
\orcid{0000-0001-7649-6493}
\affiliation{%
\institution{Shandong University}
\city{Qingdao}
 \country{China}
}
\email{zhangyunxiaox@gmail.com}

\author{Zixiong Wang}
\orcid{0000-0002-6170-7339}
\affiliation{%
 \institution{Shandong University}
 \city{Qingdao}
  \country{China}
 }
\email{zixiong_wang@outlook.com}

\author{Zihan Zhao}
\orcid{0009-0003-5962-4870}
\affiliation{%
 \institution{Shandong University}
 \city{Qingdao}
  \country{China}
 }
\email{zihanzhao2000@gmail.com}

\author{Rui Xu}
\orcid{0000-0001-8273-1808}
\affiliation{%
\institution{Shandong University}
\city{Qingdao}
 \country{China}
}
\email{xrvitd@163.com}

\author{Shuangmin Chen}
\authornote{Corresponding author}
\orcid{0000-0002-0835-3316}
\affiliation{%
 \institution{Qingdao University of Science and Technology}
 \city{Qingdao}
 \country{China}
}
\email{csmqq@163.com}

\author{Shiqing Xin}
\orcid{0000-0001-8452-8723}
\affiliation{%
\institution{Shandong University}
\city{Qingdao}
 \country{China}
}
\email{xinshiqing@sdu.edu.cn}

\author{Wenping Wang}
\orcid{0000-0002-2284-3952}
\affiliation{%
\institution{Texas A\&M University}
\city{Galveston}
 \country{USA}
}
\email{wenping@tamu.edu}

\author{Changhe Tu}
\orcid{0000-0002-1231-3392}
\affiliation{%
\institution{Shandong University}
\city{Qingdao}
 \country{China}
}
\email{chtu@sdu.edu.cn}

\renewcommand\shortauthors{Zhang, Y. et al}

\begin{abstract}
Shape manipulation is a central research topic in computer graphics. Topology editing, such as breaking apart connections, joining disconnected ends, and filling/opening a topological hole, is generally more challenging than geometry editing. 
In this paper, we observe that 
the saddle points of the signed distance function (SDF) provide useful hints for altering surface topology deliberately.
Based on this key observation, we parameterize the SDF into a cubic trivariate tensor-product B-spline function~$F$ whose saddle points~$\{\boldsymbol{s}_i\}$ can be quickly exhausted based on a subdivision-based root-finding technique coupled with Newton's method. Users can select one of the candidate points, say $\boldsymbol{s}_i$, to edit the topology in real time. In implementation, we add a compactly supported B-spline function rooted at $\boldsymbol{s}_i$, which we call a \textit{deformer} in this paper, to~$F$, with its local coordinate system aligning with the three eigenvectors of the Hessian. Combined with ray marching technique, our interactive system operates at 30 FPS. Additionally, our system empowers users to create desired bulges or concavities on the surface. An extensive user study indicates that our system is user-friendly and intuitive to operate. We demonstrate the effectiveness and usefulness of our system in a range of applications, including fixing surface reconstruction errors, artistic work design, 3D medical imaging and simulation, and antiquity restoration. Please refer to the attached video for a demonstration.

\end{abstract}

%
%



\begin{teaserfigure}
  \includegraphics[width=\textwidth]{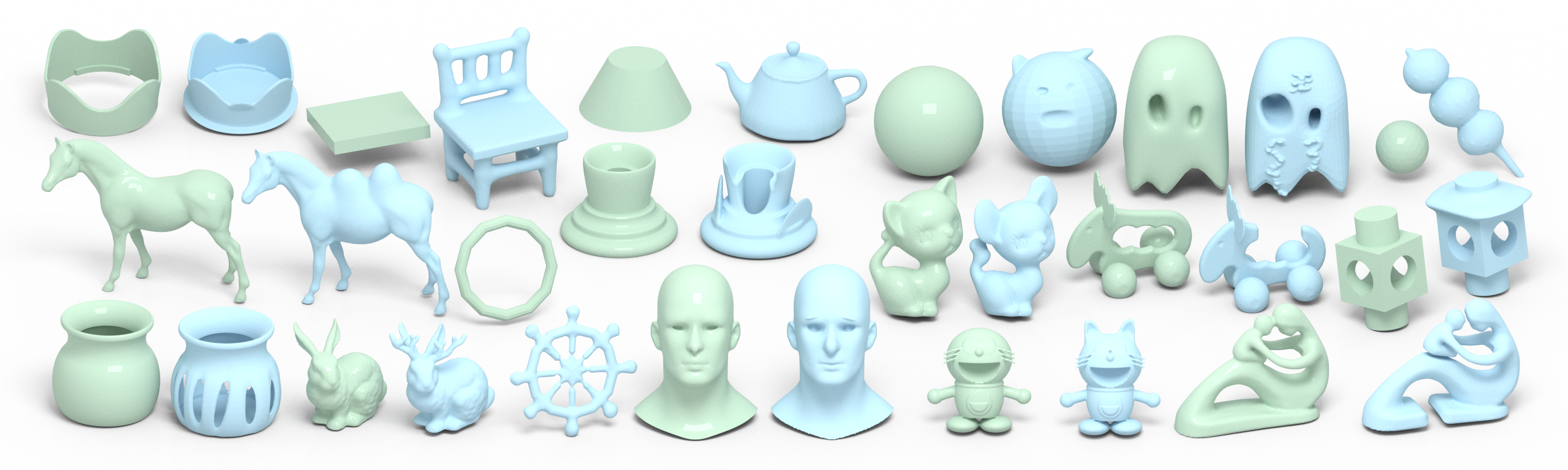}
  \vspace{-7mm}
  \caption{We present 17 groups of editing results produced by volunteers. 
  Our interactive system operates at 30 FPS, and supports both topology editing and geometry editing. 
  }
  \Description{Seventeen sets of control models, each comprising the original model and the edited model.}
  \label{fig:teaser}
\end{teaserfigure}

\maketitle

\begin{figure*}[!ht]
  \includegraphics[width=\textwidth]{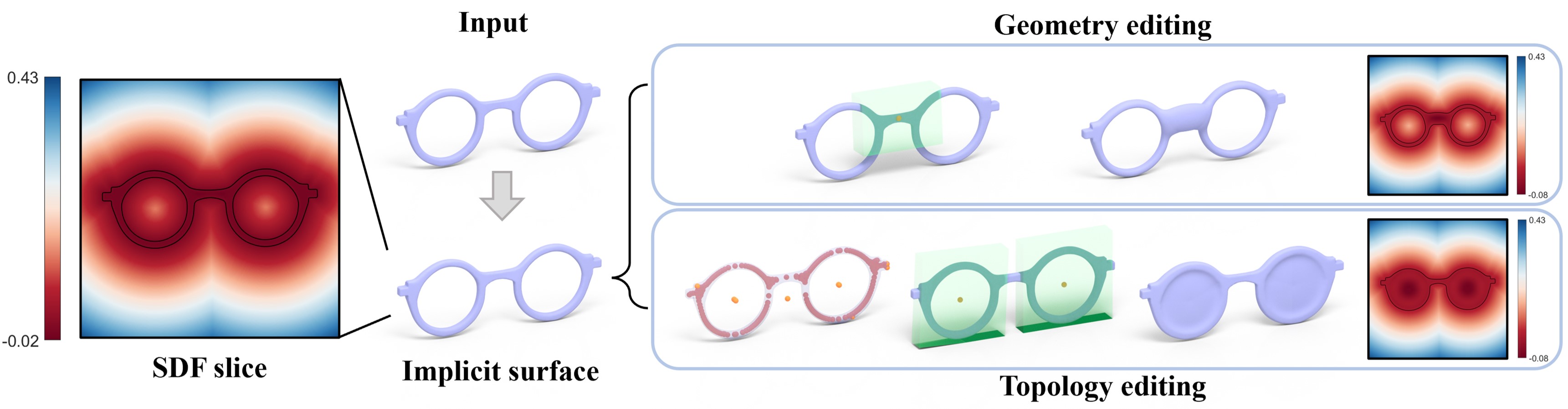}
  \vspace{-7mm}
  \caption{A typical workflow of shape manipulation with our interactive system involves operating on the implicit representation~$F$ that approximates the real signed distance function. For a base point~(in orange) on the surface, among the three eigenvectors of the Hessian of~$F$, two align with the principal curvature directions and one aligns with the normal direction. This mathematical fact enables us to define a geometry editing deformer (in green) using the local coordinate system spanned by the three eigenvectors (see the top row). Furthermore, we observe that the saddle points of~$F$ 
  are useful for determining deliberate alterations in the surface topology.
  Based on this observation, one can easily edit the topology by breaking apart connections, joining disconnected ends,
and filling or opening a topological hole~(see the bottom row).
  }
  \Description{Our paper's pipeline involves the transformation of the input model into an implicit surface, followed by topological or geometric editing to obtain the final result.}
  \vspace{-3mm}
  \label{fig:workflow}
\end{figure*}
\section{Introduction}
Shape manipulation is an active area of research in geometric modeling due to its 
ever-widening range of applications in editing tools for industrial and artistic design, such as computer-aided design~\cite{ito2021over, xiong2022algorithm}, digital sculpting~\cite{mandal2022interactive, moo2021virtual}, and character design~\cite{hu2019research, lin2022study}. 
This topic is challenging since complex mathematical formulations have to be hidden behind an intuitive user interface with which the behavior of shape manipulation can be implemented in a suﬃciently eﬃcient and robust manner.

Shape manipulation can generally be divided into two categories: topology editing and geometry editing. Topology editing allows users to modify the surface topology by breaking apart connections, joining disconnected ends, and filling or opening topological holes. Geometry editing, on the other hand, aims to manipulate surface variations. Topology editing is more challenging as it involves topological changes and is not easily performed on a triangle mesh. Some research has been done on updating implicit representations based on user-specified input strokes~\cite{angles2017sketch, ju2007editing}, but the interaction system is difficult to operate. 

In this paper, we examine the topology editing challenge through the lens of two distinct perspectives. Firstly, implicit representations such as the signed distance function~(SDF) are expressive in topology control but not intuitive for users to operate. To overcome this challenge, we parameterize the SDF into a cubic trivariate tensor-product B-spline function, which can be formulated into solving a sparse linear system. Secondly, 
if $\boldsymbol{s}$ is a saddle point of the SDF,
then one can easily change the surface topology around~$\boldsymbol{s}$
by adding a deformer rooted at~$\boldsymbol{s}$ to the SDF until
the sign of the function value at~$\boldsymbol{s}$ alters.
To be more specific, 1-saddle points, where the Hessian matrix of the SDF has only one negative eigenvalue, reveal unstable topology where one can break apart an existing connection or join two disconnected ends. 
Instead, 2-saddle points reveal unstable topology where it is easy to fill or open a topological hole. Our research is motivated by these two aspects.

In implementation, we first fit the given mesh surface into a cubic trivariate tensor-product
B-spline function~$F$, whose saddle points can be quickly exhausted by examining every cubic element. 
The saddle points enable users to select the preferred one, say $\boldsymbol{s}_i$, and edit the corresponding part. The computational task involves adding to~$F$ a compactly supported B-spline function rooted at $\boldsymbol{s}_i$, with its local coordinate system aligning with the three eigenvectors of the Hessian. Combined with ray marching technique~\cite{hart1996sphere}, our interactive system operates at 30 FPS even on an entry-level graphics processing unit.

Our contributions are three-fold:
\begin{enumerate}
\item We present a novel topology editing method 
based on the observation that the saddle points of the signed distance function (SDF) offer useful hints for determining deliberate alterations in surface topology.
\item We parameterize the SDF into a cubic trivariate tensor-product
B-spline function, allowing for quick identification of saddle points.
\item We developed a real-time interactive system that combines topology editing with geometry editing. The system runs at over 30 FPS, leveraging ray marching technique for rendering~\cite{hart1996sphere}.
An extensive user study indicates that our system is user-friendly and intuitive to operate.
\end{enumerate}

\section{Related Work}
A significant amount of research has been conducted on shape manipulation in recent decades. In this section, we review the relevant literature and categorize it based on specific representation forms.
\paragraph{Parametric/Polygonal methods}
Most existing methods manipulate shape on triangle meshes or parametric surfaces by predicting displacement for influenced mesh vertices or parametric subdomains. \citet{fowler1992geometric} and \citet{wong2000virtual} proposed a method for direct manipulation of the surface using a B-spline tensor product representation. \citet{terzopoulos1994dynamic} developed a dynamic generalization of the NURBS model for physically intuitive surface manipulation. \citet{peng2018autocomplete} advocated predicting users’ editing behavior to enable automatic editing. Despite their success in geometry editing, these surface representations are weak in topology editing because they focus on the surface itself, regardless of its embedding in three-dimensional space. Some methods~\cite{fu2004topology} consider topology requirements but involve time-consuming proximity query operations.
\paragraph{Volume-based methods}
\citet{galyean1991sculpting} first introduced volumetric representation to interactive modeling in 1991, and proposed a sculpting method based on grid-based tessellations (voxmap). 
\citet{perry2001kizamu} proposed a 3D sculpting method based on Adaptively Sampled Distance Fields (ASDF), but the approach suffers from low interaction speed. 
Benefiting from the technique for storing regular virtual grids~\cite{ferley2000practical, ferley2001resolution, barthe2002triquadratic}, \citet{baerentzen1998octree} used an octree to assist volume sculpting, resulting in higher interaction speed and less memory usage.
\citet{schmidt2005interactive} utilized hierarchical implicit volume models (blob-trees) as the underlying shape representation for interactive modeling in 2005. 
Later, some sketch-based modeling techniques~\cite{ju2007editing} are proposed 
to facilitate editing the topology of 3D models. 
The primary drawback of the aforementioned volume-based methods is that editing the volume lacks the intuitiveness of editing the surface.

\begin{figure}[!tp]
	\centering
        \begin{minipage}[!ht]{0.49\linewidth}
        \centering
        \includegraphics[width=0.95\linewidth]{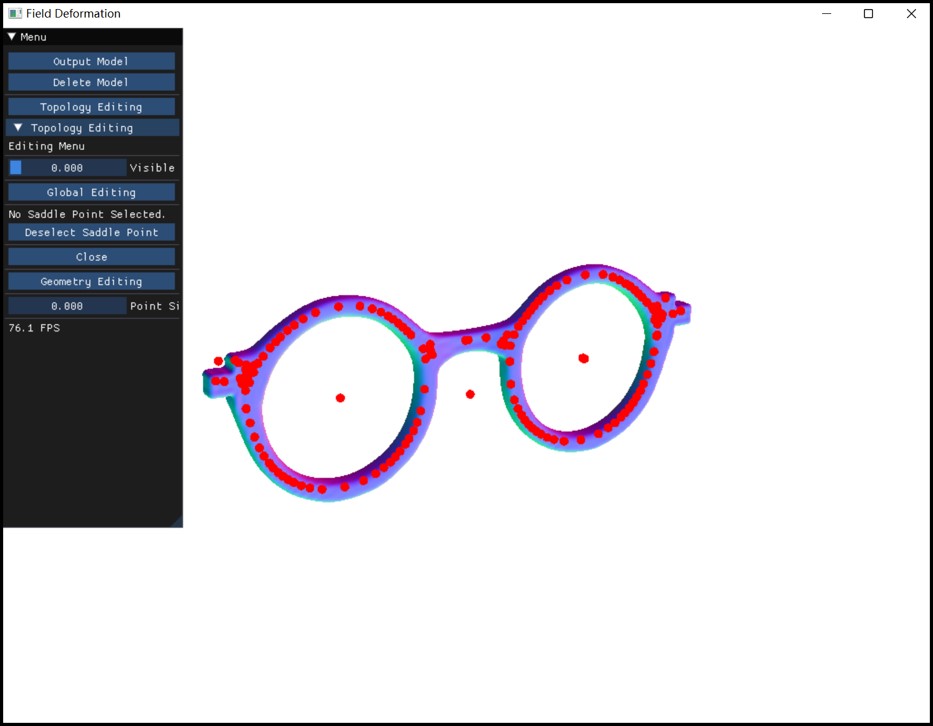}
        \end{minipage}
        \begin{minipage}[!ht]{0.49\linewidth}
        \centering
        \includegraphics[width=0.95\linewidth]{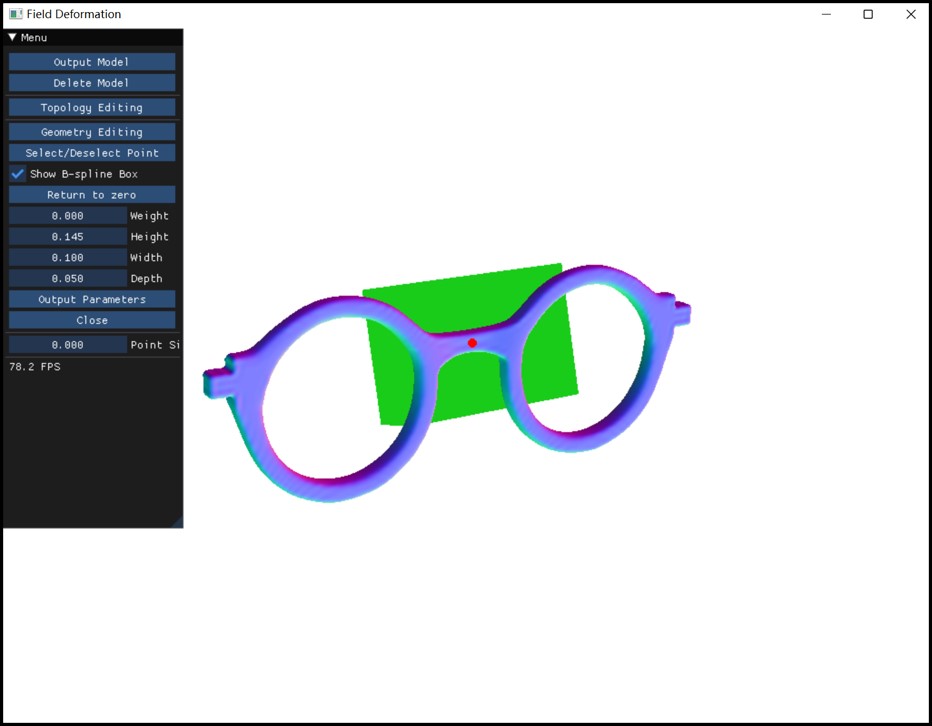}
        \end{minipage}
        \vspace{-2mm}
        \caption{
        The interface of our interactive system. 
        In the left figure, 
        we show the saddle points (in red) of the SDF. 
        In the right figure, we designate a base point where we intend to edit the geometry.
        The green box illustrates the area of influence, facilitating user adjustments.
        }
        \vspace{-8mm}
        \Description{Many red dots and a base point corresponding editing area are displayed on a pair of glasses.}
	\label{fig:ui}
\end{figure}



\paragraph{Continuous implicit function-based methods}
Implicit functions offer a flexible representation, 
enabling the definition of complex and smooth models with arbitrary topology.
\citet{hart1998morse} introduced the relationship between the topology of implicit surfaces and critical points. Building upon this relationship,
 \citet{stander1997guaranteeing} devised a polygonization algorithm that guarantees topology, based on the observation that changes in the sign of the function value at critical points lead to changes in topology, provided that the analytic form of the implicit function is given.
However, their methodology is specifically designed to extract the zero level-set polygonal surface from a given implicit function with an analytic form, rather than 
use of critical points for controlling the implicit surface topology and the related modeling tools.
\citet{sharf2007interactive} and \citet{yin2014morfit} proposed to use implicit representations 
to drive the topology change of point clouds. 
\citet{canezin2016topology} proposed a topology-aware reconstruction method.
But the run-time performance is insufficient for supporting real-time user interaction. 
\citet{museth2002level, museth2005algorithms} and \citet{eyiyurekli2010interactive} 
introduced a set of free-form editing operators based on level-set volume representations of scalar fields.
However, these methods require solving partial differential equations during each edit, and thus cannot satisfy  
interaction requirements. 
\citet{pasko2005bounded} and \citet{bernhardt2010implicit} provided a set of  blending operators that
combine two scalar functions, but this is not intuitive for users to explicitly control topology. 
\citet{gourmel2013gradient} and \citet{angles2017sketch} used gradient-based operators to manipulate shapes,
but their approach cannot be used for general shapes.
In summary, continuous implicit representations are flexible in accommodating changes in geometry and topology. However, the key challenge lies in quickly identifying the intended area for topology editing and updating the implicit representation in a manner that is easy to operate and immediately visualizes changes in topology and geometry.

\section{Hessian-based Field Deformer}
We assume that the standard input is a polygonal mesh surface. However, it's important to note that our algorithm is also compatible with other forms of surface representation. 
The workflow of our framework is outlined in Fig.~\ref{fig:workflow}, while Fig.~\ref{fig:ui} displays the interface of our interactive system. The workflow consists of three distinct steps:
1)~trivariate B-spline generation, 
2)~exhaustive search of saddle points,
and
3)~topology editing using a Hessian based deformer.
To ensure real-time interaction,
we utilize the ray marching technique~\cite{hart1996sphere} to render the 3D objects.
We will elaborate on each step in the following subsections.

\subsection{Trivariate B-Spline Generation}
Without loss of generality, 
we normalize the input model into ${[0,1]}^3$.
At the same time, we partition the unit space into 
regularly structured grid cells, where each grid vertex is
denoted by~$\boldsymbol{g}_{ijk}$. 
Let~$w$ be the distance between two adjacent grid vertices,
and $N$ be the number of grid cells along each dimension, i.e., $N=1/w$. 
All the grid vertices define a set
\begin{equation}
\boldsymbol{G}=\{\boldsymbol{g}_{ijk}:\; i,j,k=0,1,\cdots,N\}.
\end{equation}

The signed distance function (SDF),
a commonly used implicit representation,
can be readily derived from other representations, such as the point cloud~\cite{huang2019variational} and polygonal mesh~\cite{jacobson2013robust}.
Furthermore, 
it has the capability to present the topological information of any level-set in space.

Tensor-product B-splines~\cite{rouhani2014implicit, tang2018multi} have been extensively utilized to approximate the SDF. 
Owing to the local-support and non-oscillation~\cite{hall1976optimal} properties of the basis functions, we employ a cubic trivariate tensor-product B-spline function in this paper to parameterize the SDF, thereby enabling local topology editing.
The normalized cubic trivariate tensor-product B-spline basis rooted at the grid point~$\boldsymbol{g}_{ijk}$
can be written as
\begin{equation}
    B_{ijk}(\boldsymbol{q})=B(\frac{\boldsymbol{q}-\boldsymbol{g}_{ijk}}{w}). \label{eq:tensorproduct2}
\end{equation}
\begin{equation}
    B(\boldsymbol{q})=b(x)b(y)b(z),\quad\text{with}\quad \boldsymbol{q}=(x,y,z), \label{eq:tensorproduct}
\end{equation}
where $b(t)$ is the univariate cubic B-spline basis, i.e., 
\begin{equation}
    b(t)=\left\{
                \begin{array}{ll}
                  -\frac{1}{6}t^3+t^2-2t+\frac{4}{3} & t\in[1,2]\\
                  \frac{1}{2}t^3-t^2+\frac{2}{3} & t\in[0,1]\\
                  -\frac{1}{2}t^3-t^2+\frac{2}{3} & t\in[-1,0]\\
                  \frac{1}{6}t^3+t^2+2t+\frac{4}{3} & t\in[-2,-1]\\
                  0 & \mbox{otherwise}.
                \end{array}
              \right.
\end{equation}
To this end, we intend to represent the SDF in the following form
\begin{equation}
    F(\boldsymbol{q})= \sum_{i,j,k}{\alpha_{ijk}B_{ijk}(\boldsymbol{q})},\label{eq:linearsystem}
\end{equation}
where $\alpha_{ijk}$ is the unknown coefficient to be determined. 

By taking the polygonal mesh as the input, we query the SDF values
for any grid point in~$\boldsymbol{G}$ using the signed distance query routine in libigl~\cite{libigl}, and produce the following linear system:
\begin{equation}
    A\boldsymbol{\alpha}=\boldsymbol{c},\quad\text{with}\quad \boldsymbol{\alpha} = \{\alpha_{ijk}\},
\end{equation}
where  $\boldsymbol{c}$ is given by the SDF values at grid vertices, and each entry of~$A$ is computed as
\begin{equation}
    B_{ijk}(\boldsymbol{g}_{i'j'k'}),
\end{equation}
where $i,j,k$ 
traverse each basis function while $i',j',k'$ 
traverse each grid point. It is evident that $A$ is a symmetric and sparse matrix.
Furthermore, $\{B_{ijk}\}$ defines a group of basis functions that are linearly independent of each other,
from which the invertibility of~$A$ can be easily verified. 
Therefore, the coefficients $\boldsymbol{\alpha}$ can be found immediately by 
simply calling the Conjugate Gradient Method~\cite{hestenes1952methods}.
To this end, we accomplish interpolating the SDF using a cubic trivariate tensor-product B-spline function 
that enables us to analyze the first-order and second-order properties of the SDF.

\subsection{Exhaustive Search of Saddle Points}
\paragraph{Saddle points.}
$\boldsymbol{q}$ is said to be a critical point of~$F$ 
if the gradient~$\nabla F|_{\boldsymbol{q}}$ vanishes. 
As shown in Tab.~\ref{ClassifyCriticals}, the critical points of~$F$ can be further divided into 
maximum points, minimum points, 1-saddle points, and 2-saddle points,
depending on how many negative eigenvalues 
the Hessian of~$F$ owns. 
\begin{table}[!tp]
\centering
\caption{Categorization of critical points via Hessian analysis}
\Description{Critical points can be classified into minima, maxima, first-order saddle points, and second-order saddle points based on the number of positive and negative eigenvalues of their corresponding Hessian matrices.}
\begin{tabular}{l|ccc} 
\toprule
\multicolumn{1}{c|}{\multirow{2}{*}{Critical point}} & \multicolumn{3}{c}{Sign}                 \\ 
\cmidrule{2-4}
\multicolumn{1}{c|}{}                                & $\lambda_1$ & $\lambda_2$ & $\lambda_3$  \\ 
\midrule
Minimum point                                        & +           & +           & +            \\
1-saddle point                                       & -           & +           & +            \\
2-saddle point                                       & -           & -           & +            \\
Maximum point                                        & -           & -           & -            \\
\bottomrule
\end{tabular}
\vspace{-3mm}
\label{ClassifyCriticals}
\end{table}

Interestingly, a strong correlation exists between the saddle points of the SDF and the instability of the surface topology, as indicated in~\cite{stander1997guaranteeing}.
Specifically, 1-saddle points expose unstable topology where existing connections can be severed or disconnected ends can be joined. Conversely, 2-saddle points expose unstable topology where topological holes can be easily filled or opened.
\begin{figure}[!tp]
    \vspace{+2mm}
    \centering
    \includegraphics[width=\linewidth]{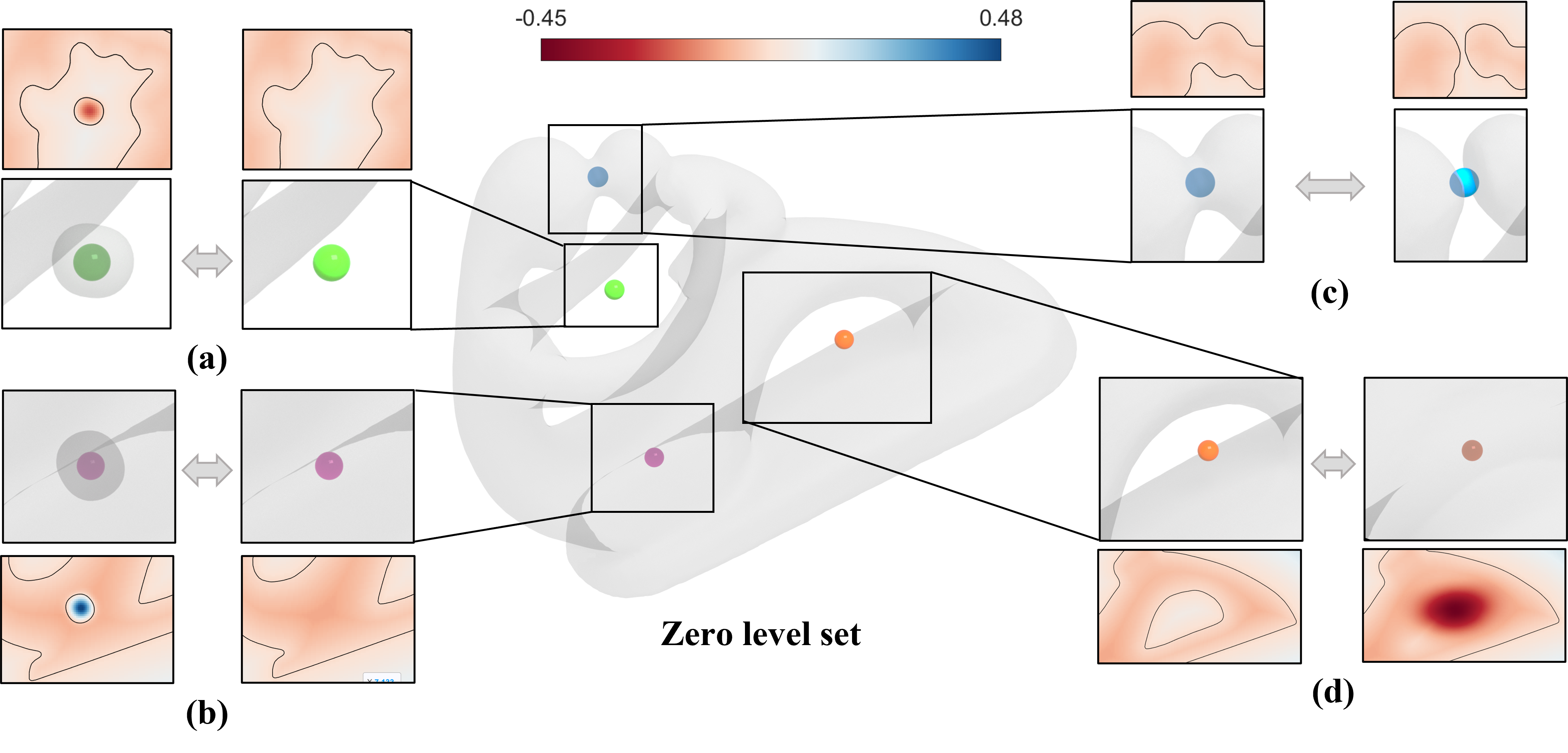}
    \vspace{-7mm}
    \caption{
    The critical points of the Signed Distance Function (SDF) are closely related to topological structures and can be classified into minimum~(purple), 1-saddle~(blue), 2-saddle~(orange), and maximum~(green) based on the Hessian of the SDF. (a-b) Maximum and minimum points reflect stable topological structures. When users make local changes to the SDF, it may produce an outlier surface that cannot be connected to the base surface. To prevent this, our system does not allow users to select maximum and minimum points for editing. (c) 1-saddle points correspond to local adhesion and separation around a thin handle. (d) 2-saddle points correspond to a tunnel area where a topological hole can be filled or opened.
    }
    \vspace{-2mm}
    \Description{For the mesh "fertility", different types of critical points correspond to distinct topological structures.}
    \label{fig:diffcriticals}
\end{figure}
Figure~\ref{fig:diffcriticals}  illustrate examples of 1-saddle, 2-saddle, maximum and minimum points, respectively.

\paragraph{Exhaustive search.}
To exhaustively search the saddle points of $F$, it is necessary to identify saddle points within each grid cell.
This can be accomplished using the subdivision-based root-finding technique~\cite{mourrain2009subdivision} to subdivide a grid cell into eight subcells.
In our implementation, we establish a maximum subdivision level of~1 to prevent the generation of an excessive number of saddle points (which could potentially be infinitely many). Subsequently, we use each subcell center as initialization and apply Newton’s method to refine the position of the saddle.
It’s important to note that the number of iterations is capped at 20.  

\subsection{Field Deformer for Topology Editing}
\begin{figure}[!tp]
    \centering
    \vspace{-2mm}
    \includegraphics[width=\linewidth]{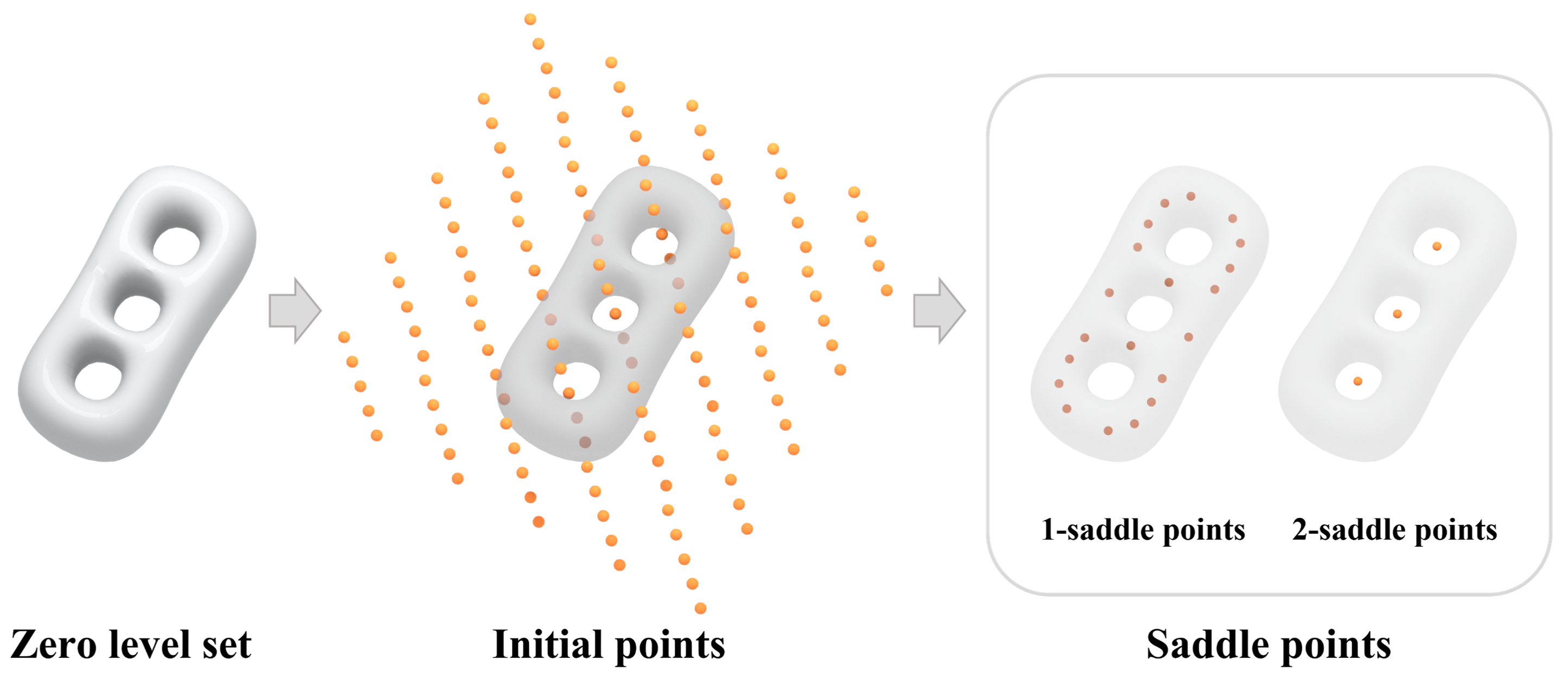}
    \vspace{-5mm}
    \caption{The process of exhaustively searching for saddle points. 
    }
    \vspace{-4mm}
    \Description{The process of exhaustively searching for critical points is divided into "Zero level set", "Initial points", "Critical points" and "Saddle points" four parts.}
    \label{fig:searchsaddles}
\end{figure}

As Figure~\ref{fig:searchsaddles} shows, 
we precompute all the saddle points~$S=\{\boldsymbol{\boldsymbol{s}_i}\}$ that indicate unstable topological structures. 
In general, users can select one of them, say $\boldsymbol{\boldsymbol{s}_i}$, and express their intention of topology editing.
The typical way for editing the implicit function~$F$ is as follows:
\begin{equation}
    F_\text{new}(\boldsymbol{q})=F(\boldsymbol{q})+\sum_{\boldsymbol{s}_i\in S}D_{\boldsymbol{\boldsymbol{s}_i}} (\boldsymbol{q}),
\end{equation}
where $D_{\boldsymbol{\boldsymbol{s}_i}} (\boldsymbol{q})$ serves as the deformer rooted at~$\boldsymbol{\boldsymbol{s}_i}$.

We suppose that the Hessian can be decomposed into
\begin{equation}
    H=Q \Lambda Q^T,
\end{equation}
where $\Lambda$ is a diagonal matrix keeping the three eigenvalues while~$Q$ encodes the three eigenvectors. 
For local shape manipulation, 
we need to align the deformer $D_{\boldsymbol{\boldsymbol{s}_i}}(\boldsymbol{q})$ with the three eigenvectors of~$H$:
\begin{equation}
    D_{\boldsymbol{\boldsymbol{s}_i}}(\boldsymbol{q})= \beta_{\boldsymbol{s}_i} B\left(W_{\boldsymbol{s}_i}^{-1}Q^T(\boldsymbol{q}-\boldsymbol{s}_i)\right),
\end{equation}
where $W_{\boldsymbol{s}_i}$ represents a diagonal matrix that defines the scaling factors along three distinct directions and $\beta_{\boldsymbol{s}_i}$ is used to tune the influence of the deformer at~$\boldsymbol{s}_i$. 

Let the saddle~$\boldsymbol{s}_i$ 
be projected onto~$\boldsymbol{s}_i'$, 
a point lying on the current surface.
Among these eigenvectors,
we take the eigenvector that closely aligns with the vector~$\boldsymbol{s}_i\boldsymbol{s}_i'$
as the first one.
Suppose that the corresponding eigenvalue is $\lambda_1$.
For the two remaining eigenvalues,
we select one of them that has the opposite sign to $\lambda_1$, and label the corresponding eigenvector as the second one. 
If both the two remaining eigenvalues own the opposite sign, 
we prioritize the one with a larger absolute value. 
To this end, the weighting parameters for the first and second eigenvectors are as follows:
$$W_{\boldsymbol{s}_i}^{(1)}=2 \left|F(\boldsymbol{s}_i)\right|,
W_{\boldsymbol{s}_i}^{(2)}=4w,$$
where $w$ is the gap between two adjacent grid points.
Finally, $W_{\boldsymbol{s}_i}^{(3)}$ is set in reference to $W_{\boldsymbol{s}_i}^{(1)}$
if their corresponding eigenvalues are of the same sign. Otherwise, $W_{\boldsymbol{s}_i}^{(3)}$ is set in reference to $W_{\boldsymbol{s}_i}^{(2)}$.
In other words, the weight ratio equals the ratio of the values of their corresponding eigenvalues.

\begin{figure}[h]
  \begin{subfigure}{\linewidth}
    \centering
    \includegraphics[width=\linewidth]{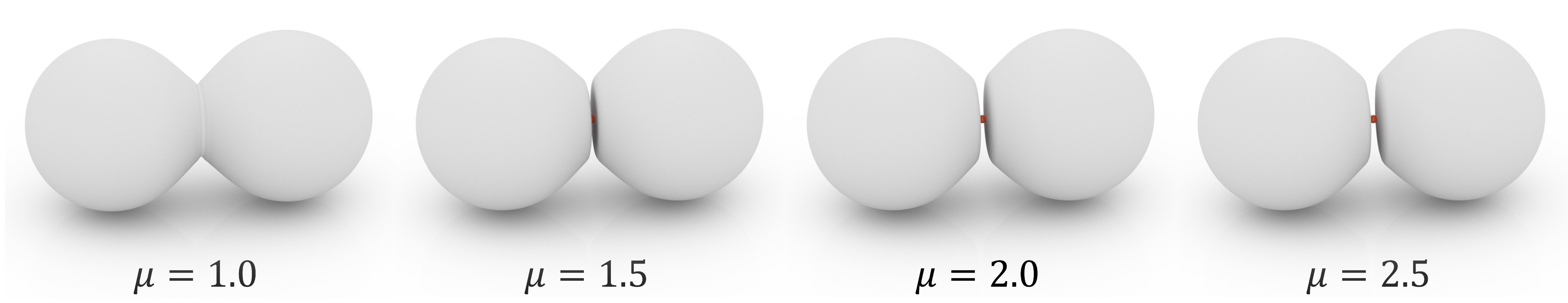}
    \caption{The effect of adjusting $\mu$, while maintaining $\varphi$ and $\rho$ at their default values.}
    \Description{When $\varphi$ and $\rho$ remain constant, the larger the value of $\mu$, the wider the area between the two spheres is separated.}
  \end{subfigure}
  \begin{subfigure}{\linewidth}
    \centering
    \includegraphics[width=\linewidth]{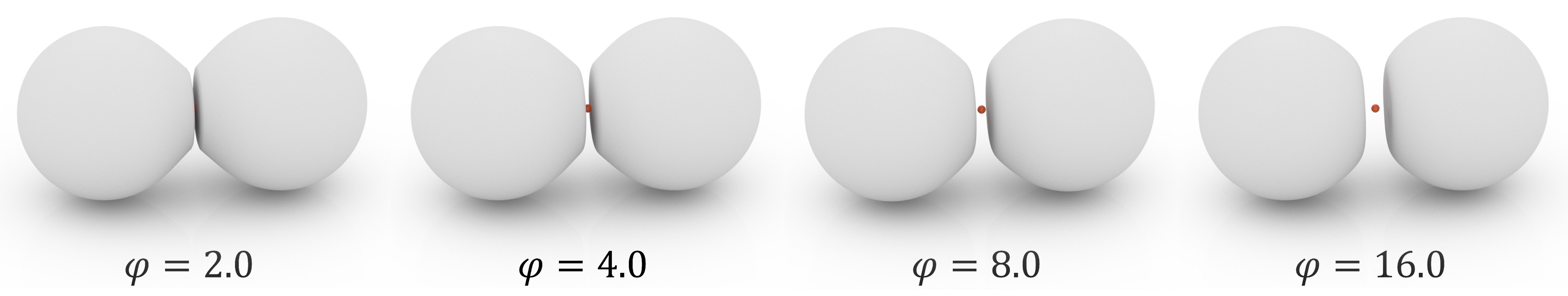}
    \caption{The effect of adjusting $\varphi$, while maintaining $\mu$ and $\rho$ at their default values.}
    \Description{When $\mu$ and $\rho$ remain unchanged, the larger the value of $\varphi$, the larger the gap between the two spheres becomes.}
  \end{subfigure}
  \begin{subfigure}{\linewidth}
    \centering
    \includegraphics[width=\linewidth]{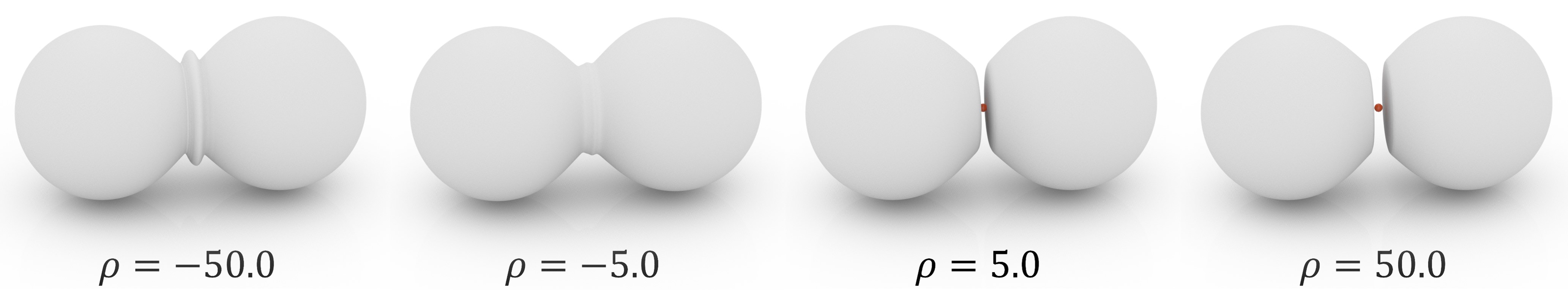}
    \caption{The effect of adjusting $\rho$, while maintaining $\mu$ and $\varphi$ at their default values.}
    \Description{When $\varphi$ and $\mu$ remain constant, the smaller the value of $\rho$, the more pronounced the adhesion effect between the two spheres, and the larger the value of $\rho$, the more pronounced the separation effect between the two spheres.}
  \end{subfigure}
  
  \caption{
  By individually adjusting the weighting parameters, one can observe distinct effects.
  We denote $W_{\boldsymbol{s}_i}^{(1)}=\mu \left|F(\boldsymbol{s}_i)\right|$ and 
  $W_{\boldsymbol{s}_i}^{(2)}=\varphi w$. The default values of $\mu$ and $\varphi$ are $2$ and $4$ respectively. 
  Additionally, $\rho=-\frac{\beta_{\boldsymbol{s}_i}}{F(\boldsymbol{s}_i)}$ 
  is used to define the overall weight of the $\boldsymbol{s}_i$-based deformer
  and set to 5 by default. }
  \Description{Different parameters have different topology editing effects on the same model.}
  \label{fig:diffweights}
\end{figure}

Under the default weight setting, 
the overall implicit function can be written as
\begin{equation}
    F_\text{new}(\boldsymbol{q})= \sum_{i,j,k}{\alpha_{ijk}B_{ijk}(\boldsymbol{q})} 
    + \sum_{\boldsymbol{s}_i\in S}\beta_{\boldsymbol{s}_i} B\left(W_{\boldsymbol{s}_i}^{-1}Q^T(\boldsymbol{q}-\boldsymbol{s}_i)\right).\label{eq:overall}
\end{equation}
To this end, users can select their preferred saddle point~$\boldsymbol{s}_i$ and subsequently modify the influence~$\beta_{\boldsymbol{s}_i}$ to generate a new implicit function, the zero level-set surface of which defines the new surface.
For example, 
if one increases $\beta_{\boldsymbol{s}_i}$ from negative to zero, $F_\text{new}(\boldsymbol{s}_i)$ 
approaches~0,
making the surface locally deformed toward the point~$\boldsymbol{s}_i$. 
It can be imagined that when one continues to increase $\beta_{\boldsymbol{s}_i}$ 
such that $F_\text{new}(\boldsymbol{s}_i)$
becomes positive, the topology will undergo a sudden change, i.e.,
\begin{equation}
    F_\text{new}(\boldsymbol{s}_i)= \sum_{i,j,k}{\alpha_{ijk}B_{ijk}(\boldsymbol{s}_i)} 
    + \beta_{\boldsymbol{s}_i} B\left(\boldsymbol{0}\right)\geq 0,\label{eq:overall_}
\end{equation}
where $B\left(\boldsymbol{0}\right)$ is $8/27$.
The above inequality implies that a topological flip between adhesion and separation may occur 
if $$\beta_{\boldsymbol{s}_i} \geq -\frac{\sum_{i,j,k}{\alpha_{ijk}B_{ijk}(\boldsymbol{s}_i)}}{B\left(\boldsymbol{0}\right)}.$$
So we set~$\beta_{\boldsymbol{s}_i} = -\rho F(\boldsymbol{s}_i)$,
where $\rho$ is a parameter. 
See Figure~\ref{fig:diffweights} for distinct effects of adjusting the weighting parameters.
\subsection{Implementation Details}
\paragraph{Geometry editing.}
In the context of topology editing, the base points for user selection are restricted to the saddle points of~$F$.
However, our approach also enables geometry editing by treating surface points as base points.
Typically, the most common geometry editing scenario involves creating a bulge or concavity.
For such a purpose, we utilize a similar 
strategy as in Section~3.3 to set the default weights for geometric editing.
The only difference lies in that
both $W_{\boldsymbol{s}_i}^{(2)}$
and $W_{\boldsymbol{s}_i}^{(3)}$
vary in reference to $W_{\boldsymbol{s}_i}^{(1)}$ 
such that the ratios between the eigenvalues are maintained. 
\begin{figure}[h]
    \centering
    \vspace{2mm}
    \includegraphics[width=\linewidth]{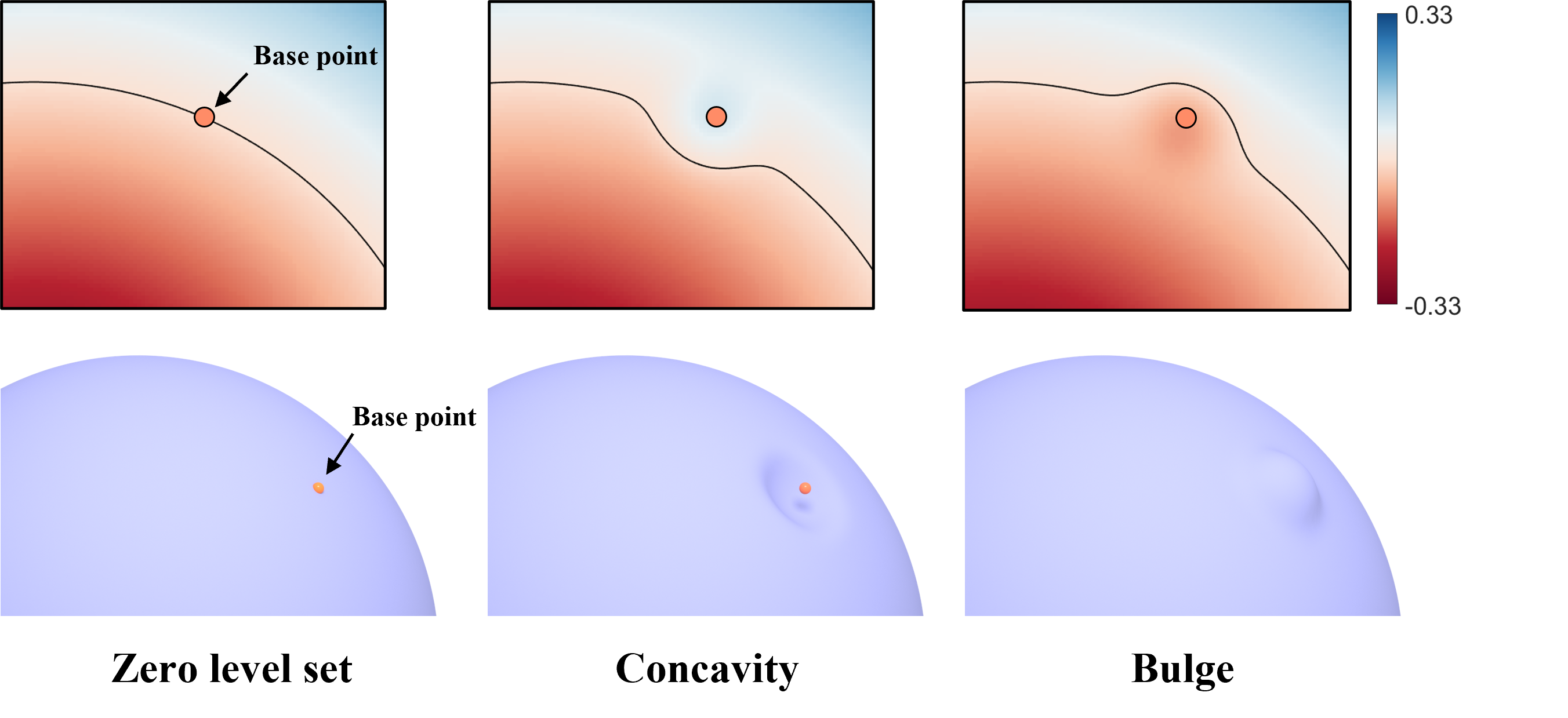}
    \vspace{-7mm}
    \caption{Two typical situations of geometry editing based on a base point~(orange).
    }
    \vspace{-1mm}
    \Description{Geometry editing can create a bulge or a concavity on the zero level set of the implicit field.}
    \label{fig:geoediting}
\end{figure}
See Figure~\ref{fig:geoediting} for an example of geometry editing.
Let $\boldsymbol{p}$ represent a point on the surface.
From differential geometry,
it is known that the Hessian matrix at a surface point~$\boldsymbol{p}$ 
must have a zero eigenvalue, with the corresponding eigenvector defining the normal vector. 
Given that~$F$ serves as an approximate counterpart of the real SDF,
we can identify the eigenvalue with the smallest absolute value and use its corresponding eigenvector as the normal vector. 
\begin{figure}[h]
    \centering
    \includegraphics[width=0.9\linewidth]{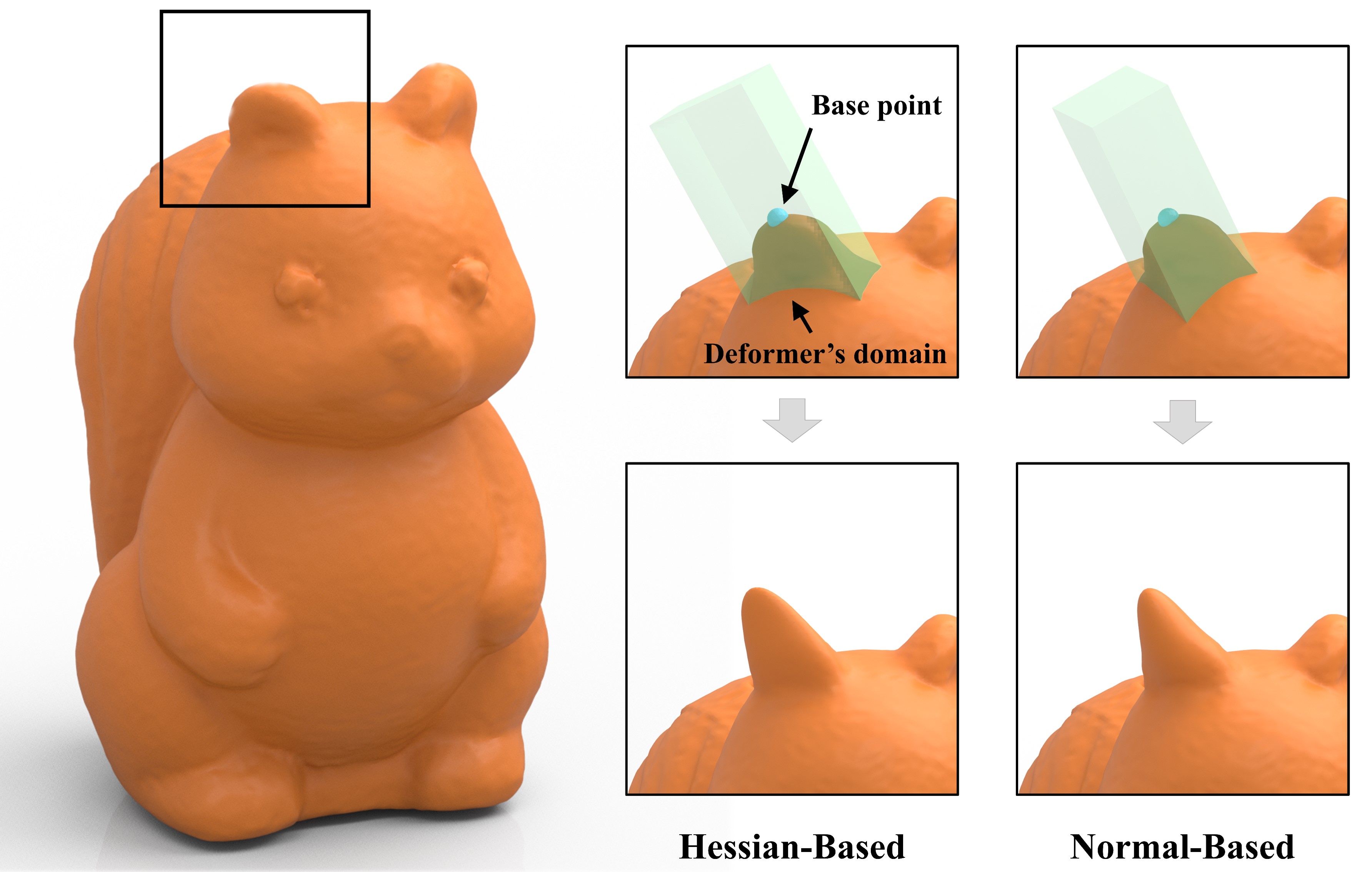}
    \caption{Our Hessian-based deformer offers a more intuitive manipulation experience compared to the normal-based deformer.}
    \Description{A round ear of the squirrel model is geometrically edited into a pointed ear using a Hessian-based deformer and normal-based deformer, respectively.}
    \label{fig:normalhessian}
\end{figure}
The other two eigenvectors naturally align with the principal curvature directions.
Our Hessian-based deformer, equipped with a carefully selected set of parameters, is capable of preserving the existing anisotropic shape variation to the greatest extent possible.
The anisotropic feature will be diminished if the eigenvector directions are replaced by a normal direction and two random orthogonal directions, which is referred to as a normal-based deformer in this paper. 
We provide a comparison between the Hessian-based deformer and the normal-based deformer in Figure~\ref{fig:normalhessian}.
The comparison suggests that our Hessian-based deformer offers a more intuitive manipulation experience compared to the normal-based deformer.

\paragraph{High frame rate.}
During the process of geometry/topology editing, users may need to adjust certain parameters to visualize the deformation outcome in real time. Clearly, marching cubes are not an ideal choice as they require visiting a large number of small cubic elements. To ensure a seamless interaction experience, we utilize the ray marching technique~\cite{hart1996sphere} to achieve a high frame rate. Additionally, it’s worth noting that the saddle points are computed only once, as minor topology editing operations do not eliminate or relocate a saddle point.
Based on this assumption, it is only necessary to retain the deformers, without the need for re-computation of the saddle points.

\section{USER STUDY}
To assess the efficiency and user-friendliness of our interactive system, which employs the proposed field deformer for topology and geometry editing, we conducted comprehensive user studies. These studies were designed to thoroughly evaluate the interactive system’s performance. We select the sketch-based topology editing method~(STEM)~\cite{ju2007editing}
as the baseline. The study comprised four stages: training, target reproduction, open-ended model creation, and feedback acquisition. All tasks were performed on a laptop equipped with an Intel Core i7-12700H Processor, 16GB of memory, and an NVIDIA GeForce RTX 3060 graphics processing unit.


\subsection{Participants and User Training}


We enlisted 15 volunteers, comprising 3 females and 12 males, aged between 18 and 30. The volunteers were undergraduate and graduate students with backgrounds in computer science education. Of these volunteers, eight have prior experience with model editing systems, while the remaining seven are novices.

We trained the participants by explaining the concepts of topology and geometry editing and introducing the tools, including our system and STEM~\cite{ju2007editing}, over a period of 5 minutes. To ensure that all participants were proficient in using the test tools, we asked them to complete several simple training tasks similar to our reproduction tasks.

\subsection{Target Reproduction Tasks}
To thoroughly demonstrate our system’s performance, we meticulously designed four representative tasks that focus on different types of topology editing: opening/filling a topological hole, breaking apart connections, and joining disconnected ends.

\paragraph{Task models construction}

To ensure the suitability of the input/target models pair for the tasks, we manually created the input model of each task based on a selected target models using boolean operators in MeshLab~\cite{meshlab}. The model is presented in the first column of Figure~\ref{fig:task} (on Page 9). 

\paragraph{Reproduction procedure}
For each task, participants were provided with an input model and a reference target model. For comparison purposes, they were instructed to edit the input model to closely resemble the target model using both STEM~\cite{ju2007editing} and our system. Each task had a maximum time limit of 5 minutes, and participants were asked to fill in a questionnaire. Additionally, the quality of the results was evaluated using common metrics.


\paragraph{Quantitative evaluation}
We employ a combination of objective and subjective methods to quantitatively evaluate both the results and user experience.

Objectively, in order to evaluate the quality of the results in target reproduction tasks, we selected the following metrics: chamfer distance (scaled by $10^3$, using $L_1$-norm) measures the fitting tightness between the two surfaces; F-Score (\%) indicates the harmonic mean of precision and recall (completeness); normal consistency (\%, abbreviated as
‘Normal C.’) reflects the degree to which the normals of the edited surface agree with the normals of the ground-truth surface; genus and connected components present topological measures over the edited models. 
As presented in Table~\ref{tab:quan_topology}~(on Page 10), our system outperforms STEM~\cite{ju2007editing} in all metrics, showing that users achieved better-quality outcomes by our system within the same editing time. 
Compared to STEM~\cite{ju2007editing} in which user-drawn sketches may not accurately convey the intended interaction, our system utilizes precomputed saddle points as interaction hints and offers a more straightforward and intuitive manipulation method.

\begin{figure}[H]
  \centering
\includegraphics[width=0.9\linewidth]{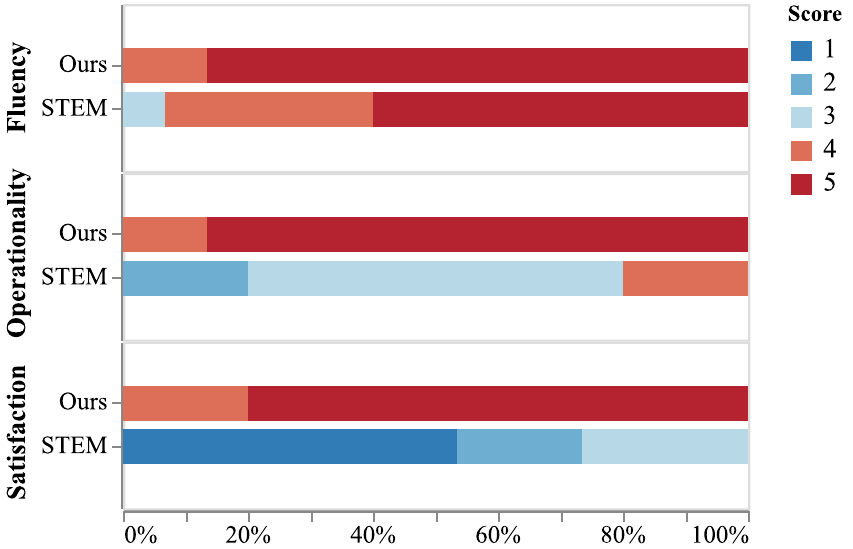}
\vspace{-2mm}
\caption{Statistics about fluency, operationality and satisfaction
based on the questionnaire.}
\Description{Our system received high scores on system fluency, system operationality, and satisfaction with the editing results, while STEM has poor operationality and most participants were not satisfied with the editing results.}
\vspace{-1mm}
 \label{fig:rate_results}
\end{figure}

Subjectively, all participants filled in the questionnaire including three scoring items: system fluency, system operationality, and satisfaction with the editing results on a scale of~1 to~5. The statistical results are shown in Figure~\ref{fig:rate_results} (on Page 9). It shows that our system received high scores on all items, while STEM has poor operationality and most participants were not satisfied with the editing results.









\subsection{Open-ended Model Creation and User Feedbacks}
\begin{figure}[H]
    \centering
    \includegraphics[width=\linewidth]{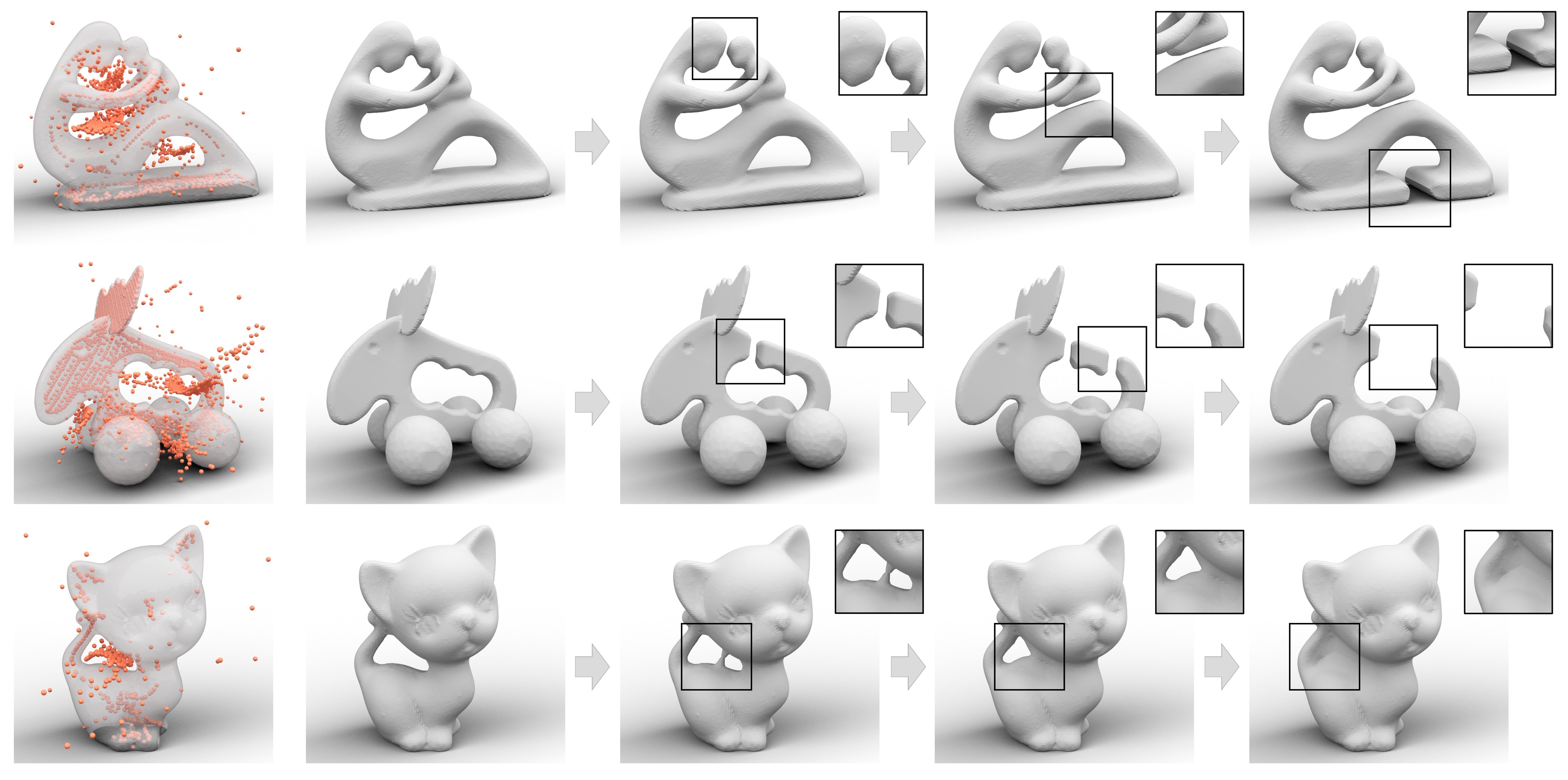}
    \caption{Examples of the creative process exclusively achieved through topology editing. The entire process comprises two parts: saddle point localization (orange balls in leftmost column) and interactive editing (right four columns).}
    \Description{The saddle points of the three models and the entire process of editing their topology.}
    \label{fig:freetopoediting}
\end{figure}

\begin{figure}[H]
    \centering
    \includegraphics[width=\linewidth]{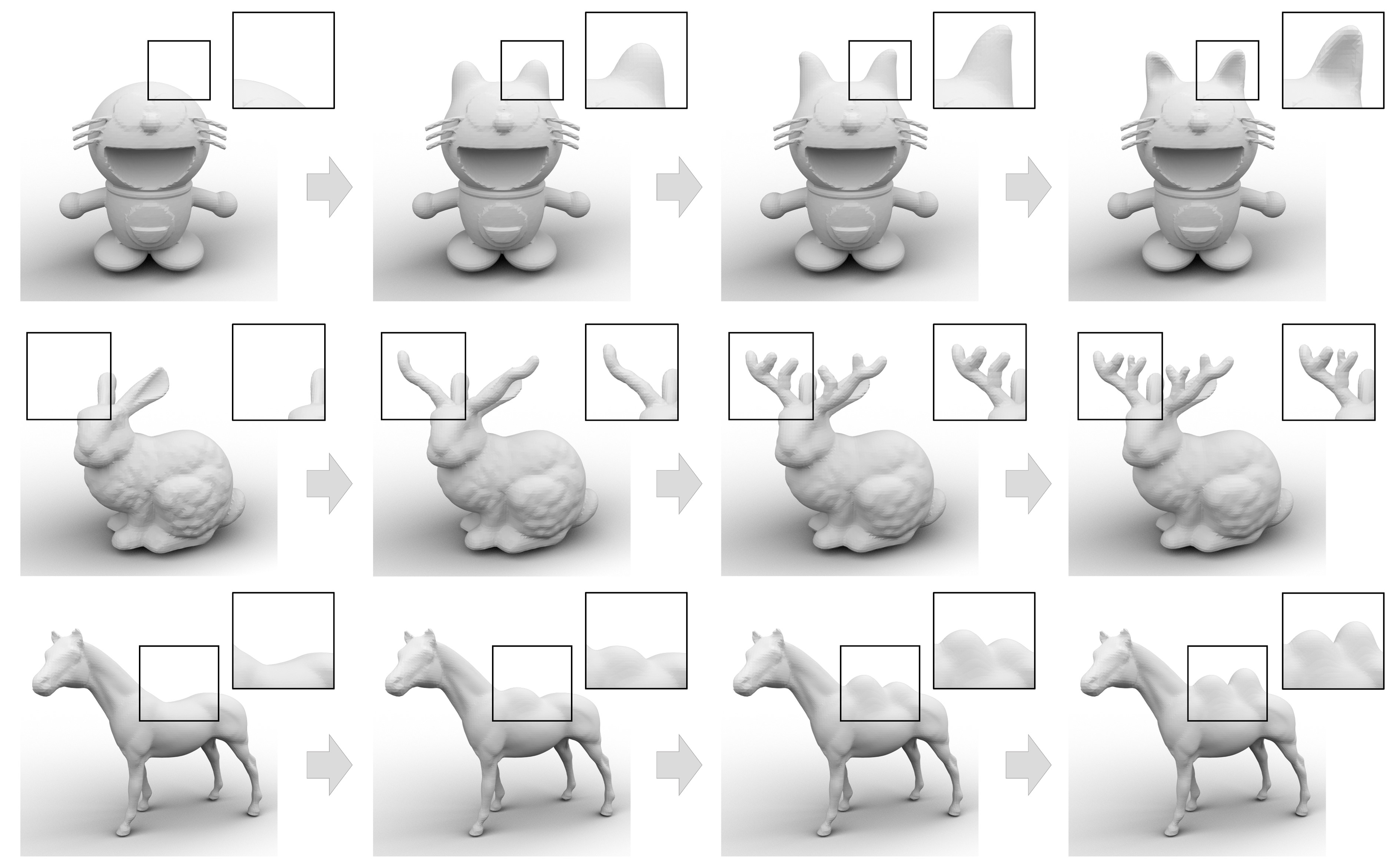}
    \caption{Examples of the creative process exclusively achieved through geometry editing.}
    \Description{The process of modifying the geometric features of three models.}
    \label{fig:freegeoediting}
\end{figure}

To evaluate the creative freedom provided by our system, all participants were 
allowed to make topological and geometric changes.
A selection of the results is presented in Figure~\ref{fig:teaser}.
In Figure~\ref{fig:freetopoediting} and Figure~\ref{fig:freegeoediting} (both on Page 9), we present examples created through either topology editing or geometry editing, respectively.
Upon the completion of all tasks, we conducted brief interviews with the participants to get their impressions of our system. The majority of users reported that the system was straightforward to learn and significantly reduced the complexity and learning curve associated with model editing. One beginner user remarked, ``Compared to other commercial tools, this system simplifies the learning process for 3D creation. I can now effortlessly create the models I desire.''


\subsection{Performance}
We collected necessary statistics to demonstrate the efficiency of our system.

\begin{figure}[H]
    \centering
    \includegraphics[width=\linewidth]{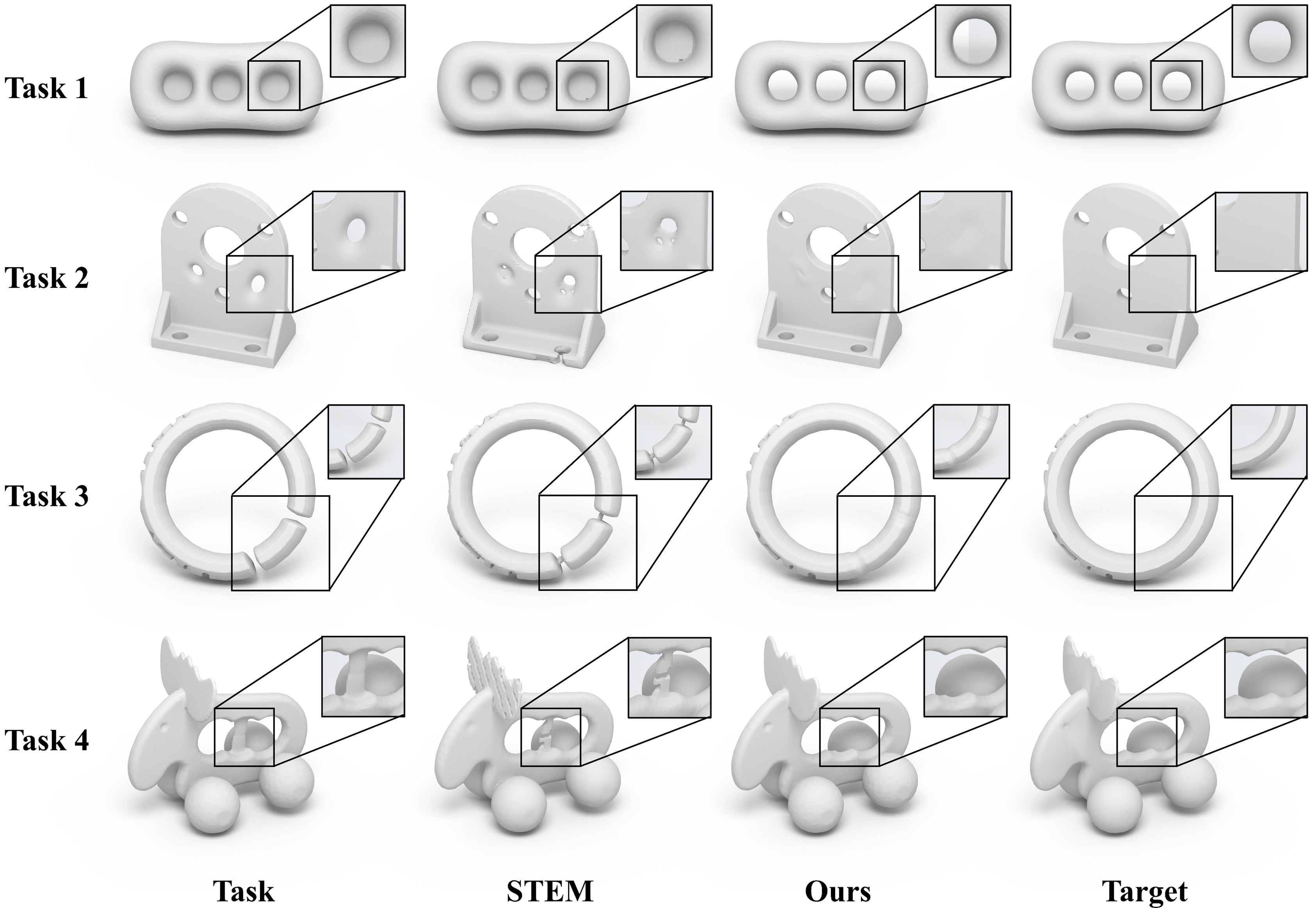}
    \caption{Test our system and STEM~\cite{ju2007editing} 
    by conducting four target reproduction tasks. It can be seen from the highlighted differences that our system is capable of more effectively restoring real topological and geometric structures.}
    \Description{Our method excels at accurately restoring the topology and geometric features of the target model, whereas the STEM falls short in this regard.
    }
    \label{fig:task}
\end{figure}

\begin{figure}[H]
  \centering
   \includegraphics[width=0.9\linewidth]{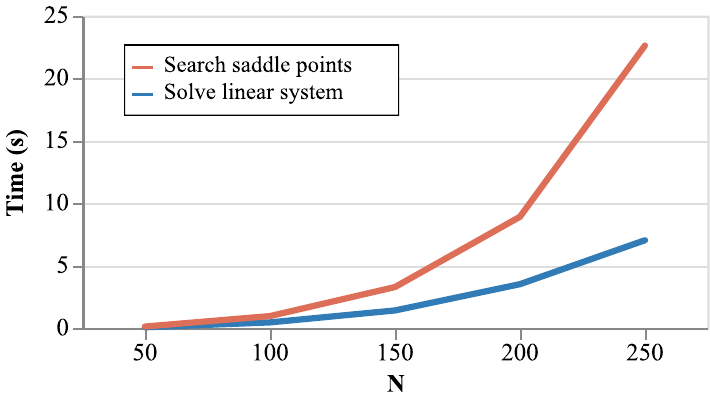}
 \vspace{-2mm}
 \caption{Timing costs with regard to the grid resolution along an axis.}
 \Description{The computering time of searching saddle points and solving linear system increases with the increase of $N$. Among them, the former increases faster.}
 \vspace{-1mm}
 \label{fig:performance_search_solve}
\end{figure}

\paragraph{Time cost for implicitization and exhaustive search of saddles.} 
Since our algorithm operates directly on the tensor-product function,
the resolution of grid cells plays a crucial role in balancing the accuracy and run-time performance of implicitization and exhaustive search of saddles. The relationship between resolution and run-time costs is illustrated in Figure~\ref{fig:performance_search_solve} (on Page 9). 
In our user study, we used a resolution of $N=150$, as shown in Figure~\ref{fig:task} (on Page 9). With this setting, the search cost is typically below 25 seconds and the time required to solve equations is reduced to less than 10 seconds.


\paragraph{Interactive speed} 
To evaluate the real-time performance of our system, we measured the Frames Per Second (FPS) for topology editing, geometry editing, and idle state over a 5-minute duration using two different graphics processing units: the NVIDIA GeForce GTX 1060 and NVIDIA GeForce RTX 3060, based on the same CPU. As shown in Figure~\ref{fig:performance_fps} (on Page 9), our system operates at above 30 FPS on whether 1060 or 3060, providing a high frame rate and enabling real-time response to user operations.



\section{Applications}
We present several example applications that benefit from the Hessian-based field deformer. In different application scenarios, users have the flexibility to selectively choose either saddle points or any point on the model surface based on their specific requirements. 

\paragraph{Fixing surface reconstruction errors}
Reconstructing a continuous surface from a given point cloud has always been a highly challenging task, particularly in cases where the point cloud exhibits poor quality, such as sparsity, noise, and other defects. Screened Poisson reconstruction~(SPR)~\cite{kazhdan2013screened}, one of the most classical reconstruction methods, also faces the challenge of introducing topological noise when reconstructing from low-quality point clouds. In fact, topological errors such as unexpected adhesion, separation, and holes may occur, greatly limiting its use in many practical applications. To our knowledge, the research community lacks an easy-to-use tool to fix these issues on reconstruction results. With the support of our Hessian-based field deformer, one can easily edit geometry variations or topological errors with minimal computational effort. In general, users can use our default parameters for quick geometry/topology editing or customize a set of new parameters for specific application scenarios. Figure~\ref{fig:app1}~(on Page 10) illustrates several outcomes achieved by utilizing our tool to fix issues caused by SPR. It is evident that our deformers can easily recover the original topological structure without altering topologically faithful areas.

\begin{figure}[H]
    \centering
    \includegraphics[width=\linewidth]{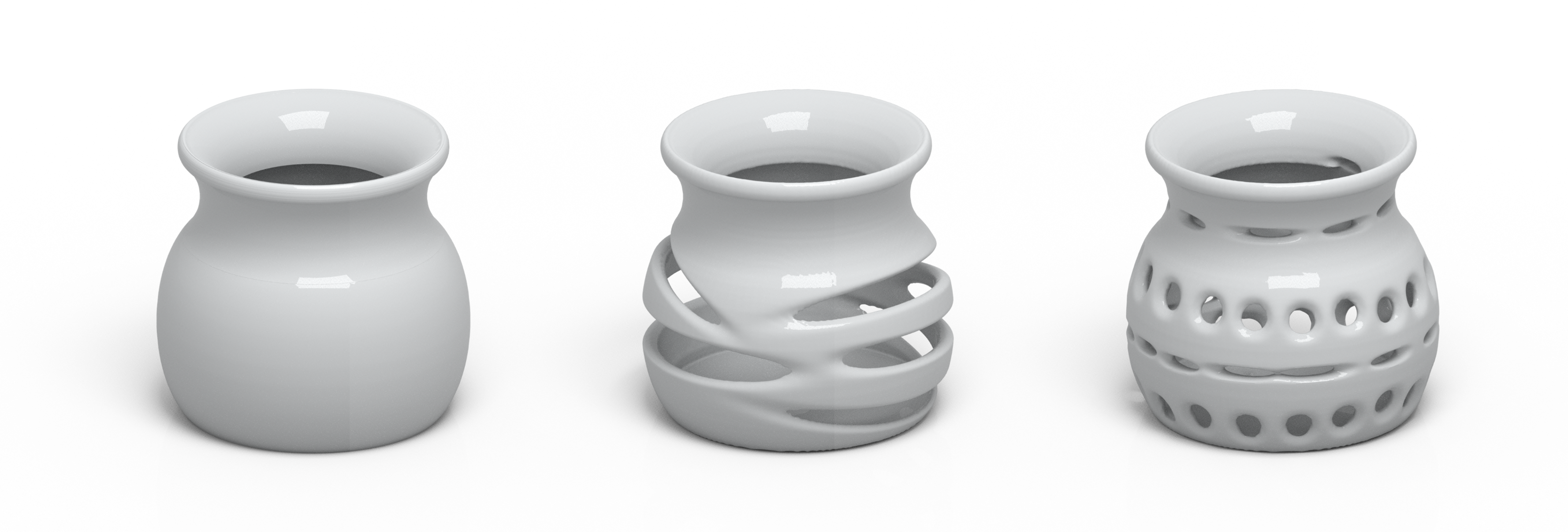}
    \caption{
    Taking the short honey pot (left) as the base model,
    we allow users to edit the topology according to their own wishes,
    yielding two creative artistic designs (middle, right).
    }
    \Description{Three honey pots. Two of them were obtained through topology editing by the other one.}
    \label{fig:app2}
\end{figure}

\paragraph{Artistic design}
Designing a work of art is often a complex and intricate task. During the design process, artists need to continuously modify the topology and geometry of the model to achieve artistic and aesthetic appeal.
However, existing design tools such as ZBrush, Blender, etc., struggle with direct topological modifications, which greatly hinders the design of art pieces. Our Hessian-based field deformer, on the other hand, has the ability to directly change the topology of the model based on saddle points in an intuitive manner, providing artists with tremendous convenience. 
During the actual editing process, artists can select the saddle points corresponding to the area of interest to flip the local topology toward the desired structure.
Refer to Figure~\ref{fig:app2}~(on Page 10) for an illustration of the results achieved using our deformer.

\begin{figure}[H]
    \centering
    \includegraphics[width=\linewidth]{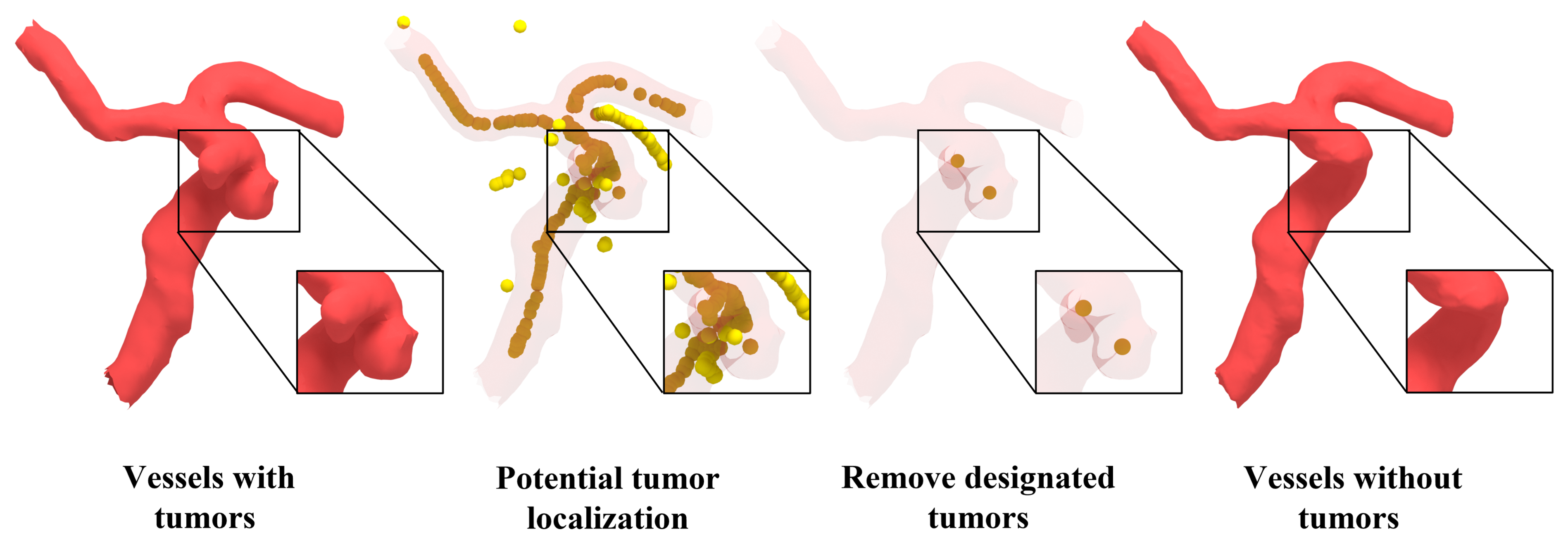}
    \vspace{-2mm}
    \caption{With the use of our interactive system, doctors are able to quickly locate vascular tumors and remove them from the original model, resulting in a virtually healed vessel model.    
    }
    \Description{The left image shows a partial mesh of the vascular system. The middle image displays all the located saddle points. The right image depicts the vascular system after simulating the removal of the vascular tumor.}
    \label{fig:app3}
\end{figure}

\paragraph{3D medical imaging/simulation}
Our method also finds application in medical imaging and simulation, allowing for the manipulation of the topology of anatomical models. This enables customized surgical planning, simulation, and educational purposes, enhancing patient-specific treatment strategies and surgical outcomes. By utilizing the Hessian-based field deformer, surgeons and researchers can interactively edit the topology of anatomical models, visualize anatomical structures, identify surgical risks, and optimize surgical strategies. For example, due to the unique topological structure of vascular tumors within the vascular system, our interactive system can rapidly and accurately locate potential vascular tumors and simulate their removal, as depicted in Figure~\ref{fig:app3}~(on Page 10).

\begin{figure}[H]
    \centering
    \includegraphics[width=\linewidth]{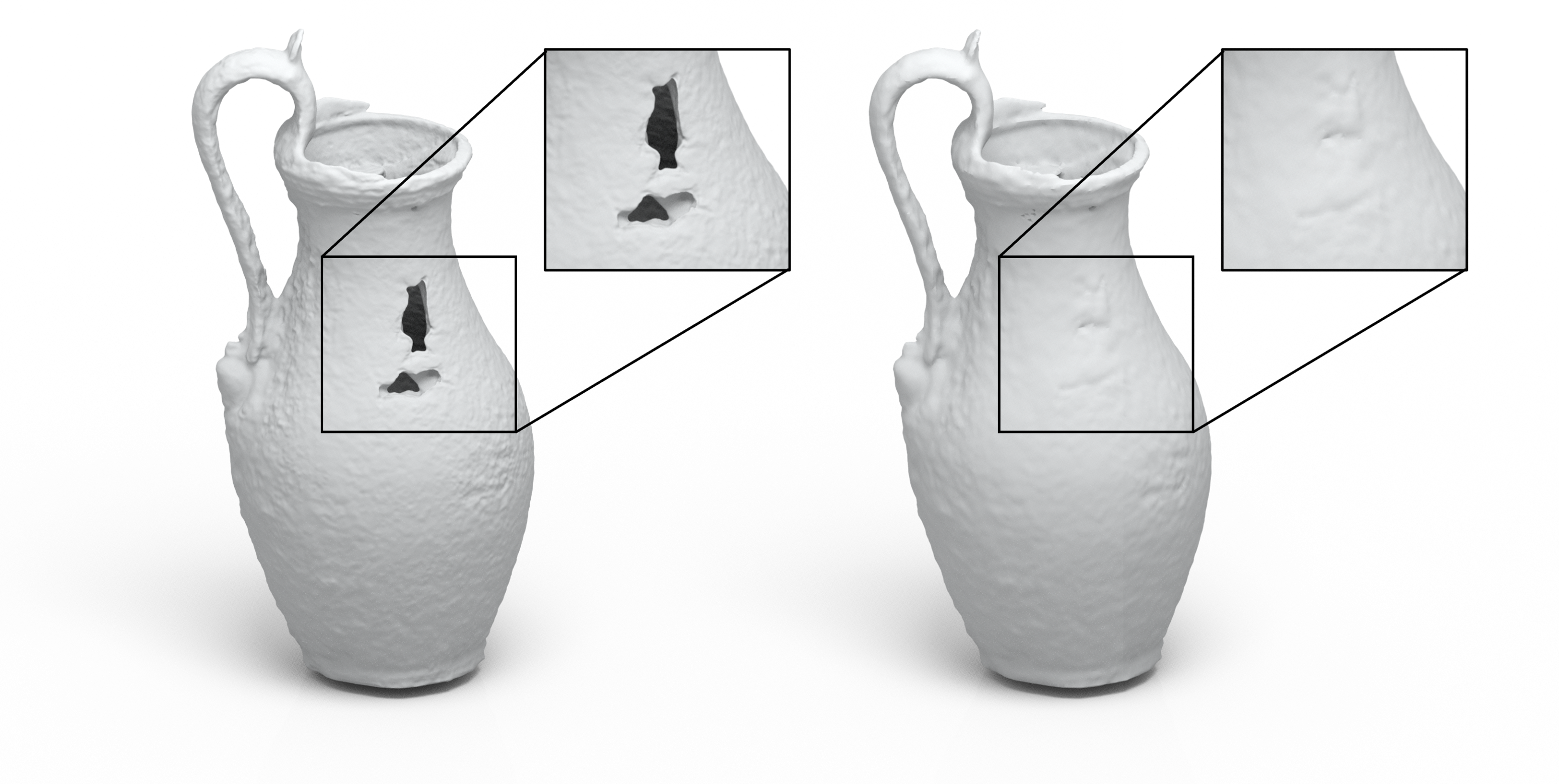}
    \caption{The restoration outcome of an ancient Roman wine-jug by utilizing our interactive system.
    }
    \Description{The wine-jug on the left has a hole in the body. The hole on the body of the wine-jug on the right has been filled with our method.}
    \label{fig:app4}
\end{figure}

\paragraph{Antiquity restoration}
Another application of our method involves cultural heritage restoration, specifically in the domain of antiquity conservation. These antiquities possess significant historical and cultural value. In the event of irreversible damage, physical restoration can be laborious and protracted, potentially compromising their intrinsic value. Instead, our system offers a convenient tool to identify structural issues and edit the geometry/topology within the virtual world.
Figure~\ref{fig:app4}~(on Page 10) showcases an example of ancient antiquity restoration by utilizing using our system.
It can be seen that our system enables professionals to achieve their intentional modifications in an intuitive manner.
To sum up, our interactive system can help professionals enhance their efficiency and improve restoration outcomes.


\section{Conclusions}
We present a Hessian-based field deformer that facilitates real-time topology-aware shape editing, guided by two primary considerations.
First, a strong correlation exists between the saddle points of the SDF and the unstable topological configurations of the surface. Second, by parameterizing the SDF as a cubic trivariate tensor-product B-spline function, we construct the implicitization with second-order smoothness and local control properties, enabling the rapid identification of all saddle points.
Utilizing ray marching technique~\cite{hart1996sphere},
 we developed an interactive system for topology and geometry editing that is user-friendly and intuitive to operate based on extensive user studies.
We further demonstrate the effectiveness and usefulness of our system in a range of applications, including fixing surface reconstruction errors, artistic work design, 3D medical imaging and simulation, and virtual antiquity restoration.


\paragraph{Limitations}
Our methodology, in its present form, has at least two limitations. First, it operates directly on the implicit representation, which can lead to slight deviations from the original surface due to the implicitization process. This issue is particularly pronounced for CAD models.
Second, our current implementation relies on uniform space partitioning, resulting in increased time and memory overhead for high-resolution scenarios.



\paragraph{Future works}
We intend to define additional deformer templates to accommodate a wider variety of editing effects.
Furthermore, we will add more buttons to the interactive system, such as copy, paste, undo, and redo, to improve the user experience.
Additionally, we shall improve the implementation, specifically by exploring the potential of substituting the regular grid with an octree.
\begin{acks}
The authors would like to thank the anonymous reviewers for their valuable comments and suggestions. This work is supported by National Key R\&D Program of China (2022YFB3303200), National Natural Science Foundation of China (62002190, 62272277), and Natural Science Foundation of Shandong Province (ZR2020MF036).
\end{acks}

\bibliographystyle{ACM-Reference-Format}
\bibliography{reference}


\begin{thebibliography}{44}


\ifx \showCODEN    \undefined \def \showCODEN     #1{\unskip}     \fi
\ifx \showDOI      \undefined \def \showDOI       #1{#1}\fi
\ifx \showISBNx    \undefined \def \showISBNx     #1{\unskip}     \fi
\ifx \showISBNxiii \undefined \def \showISBNxiii  #1{\unskip}     \fi
\ifx \showISSN     \undefined \def \showISSN      #1{\unskip}     \fi
\ifx \showLCCN     \undefined \def \showLCCN      #1{\unskip}     \fi
\ifx \shownote     \undefined \def \shownote      #1{#1}          \fi
\ifx \showarticletitle \undefined \def \showarticletitle #1{#1}   \fi
\ifx \showURL      \undefined \def \showURL       {\relax}        \fi
\providecommand\bibfield[2]{#2}
\providecommand\bibinfo[2]{#2}
\providecommand\natexlab[1]{#1}
\providecommand\showeprint[2][]{arXiv:#2}

\bibitem[Angles et~al\mbox{.}(2017)]%
        {angles2017sketch}
\bibfield{author}{\bibinfo{person}{Baptiste Angles}, \bibinfo{person}{Marco Tarini}, \bibinfo{person}{Brian Wyvill}, \bibinfo{person}{Lo{\"\i}c Barthe}, {and} \bibinfo{person}{Andrea Tagliasacchi}.} \bibinfo{year}{2017}\natexlab{}.
\newblock \showarticletitle{Sketch-based implicit blending}.
\newblock \bibinfo{journal}{\emph{ACM Transactions on Graphics (TOG)}} \bibinfo{volume}{36}, \bibinfo{number}{6} (\bibinfo{year}{2017}), \bibinfo{pages}{1--13}.
\newblock


\bibitem[B{\ae}rentzen(1998)]%
        {baerentzen1998octree}
\bibfield{author}{\bibinfo{person}{Andreas B{\ae}rentzen}.} \bibinfo{year}{1998}\natexlab{}.
\newblock \showarticletitle{Octree-based volume sculpting}. In \bibinfo{booktitle}{\emph{IEEE Visualization}}, Vol.~\bibinfo{volume}{98}. \bibinfo{pages}{9--12}.
\newblock


\bibitem[Barthe et~al\mbox{.}(2002)]%
        {barthe2002triquadratic}
\bibfield{author}{\bibinfo{person}{Lo{\"\i}c Barthe}, \bibinfo{person}{Benjamin Mora}, \bibinfo{person}{Neil Dodgson}, {and} \bibinfo{person}{Malcolm Sabin}.} \bibinfo{year}{2002}\natexlab{}.
\newblock \showarticletitle{Triquadratic reconstruction for interactive modelling of potential fields}. In \bibinfo{booktitle}{\emph{Proceedings SMI. Shape Modeling International 2002}}. IEEE, \bibinfo{pages}{145--275}.
\newblock


\bibitem[Bernhardt et~al\mbox{.}(2010)]%
        {bernhardt2010implicit}
\bibfield{author}{\bibinfo{person}{Adrien Bernhardt}, \bibinfo{person}{Loic Barthe}, \bibinfo{person}{Marie-Paule Cani}, {and} \bibinfo{person}{Brian Wyvill}.} \bibinfo{year}{2010}\natexlab{}.
\newblock \showarticletitle{Implicit blending revisited}. In \bibinfo{booktitle}{\emph{Computer Graphics Forum}}, Vol.~\bibinfo{volume}{29}. Wiley Online Library, \bibinfo{pages}{367--375}.
\newblock


\bibitem[Canezin et~al\mbox{.}(2016)]%
        {canezin2016topology}
\bibfield{author}{\bibinfo{person}{Florian Canezin}, \bibinfo{person}{Ga{\"e}l Guennebaud}, {and} \bibinfo{person}{Lo{\"\i}c Barthe}.} \bibinfo{year}{2016}\natexlab{}.
\newblock \showarticletitle{Topology-aware neighborhoods for point-based simulation and reconstruction}. In \bibinfo{booktitle}{\emph{Eurographics/ACM SIGGRAPH Symposium on Computer Animation}}.
\newblock


\bibitem[Cignoni et~al\mbox{.}(2008)]%
        {meshlab}
\bibfield{author}{\bibinfo{person}{Paolo Cignoni}, \bibinfo{person}{Marco Callieri}, \bibinfo{person}{Massimiliano Corsini}, \bibinfo{person}{Matteo Dellepiane}, \bibinfo{person}{Fabio Ganovelli}, {and} \bibinfo{person}{Guido Ranzuglia}.} \bibinfo{year}{2008}\natexlab{}.
\newblock \showarticletitle{{MeshLab: an Open-Source Mesh Processing Tool}}. In \bibinfo{booktitle}{\emph{Eurographics Italian Chapter Conference}}, \bibfield{editor}{\bibinfo{person}{Vittorio Scarano}, \bibinfo{person}{Rosario~De Chiara}, {and} \bibinfo{person}{Ugo Erra}} (Eds.). \bibinfo{publisher}{The Eurographics Association}.
\newblock


\bibitem[Eyiyurekli and Breen(2010)]%
        {eyiyurekli2010interactive}
\bibfield{author}{\bibinfo{person}{Manolya Eyiyurekli} {and} \bibinfo{person}{David Breen}.} \bibinfo{year}{2010}\natexlab{}.
\newblock \showarticletitle{Interactive free-form level-set surface-editing operators}.
\newblock \bibinfo{journal}{\emph{Computers \& Graphics}} \bibinfo{volume}{34}, \bibinfo{number}{5} (\bibinfo{year}{2010}), \bibinfo{pages}{621--638}.
\newblock


\bibitem[Ferley et~al\mbox{.}(2000)]%
        {ferley2000practical}
\bibfield{author}{\bibinfo{person}{Eric Ferley}, \bibinfo{person}{Marie-Paule Cani}, {and} \bibinfo{person}{Jean-Dominique Gascuel}.} \bibinfo{year}{2000}\natexlab{}.
\newblock \showarticletitle{Practical volumetric sculpting}.
\newblock \bibinfo{journal}{\emph{The Visual Computer}} \bibinfo{volume}{16}, \bibinfo{number}{8} (\bibinfo{year}{2000}), \bibinfo{pages}{469--480}.
\newblock


\bibitem[Ferley et~al\mbox{.}(2001)]%
        {ferley2001resolution}
\bibfield{author}{\bibinfo{person}{Eric Ferley}, \bibinfo{person}{Marie-Paule Cani}, {and} \bibinfo{person}{Jean-Dominique Gascuel}.} \bibinfo{year}{2001}\natexlab{}.
\newblock \showarticletitle{Resolution adaptive volume sculpting}.
\newblock \bibinfo{journal}{\emph{Graphical Models}} \bibinfo{volume}{63}, \bibinfo{number}{6} (\bibinfo{year}{2001}), \bibinfo{pages}{459--478}.
\newblock


\bibitem[Fowler(1992)]%
        {fowler1992geometric}
\bibfield{author}{\bibinfo{person}{Barry Fowler}.} \bibinfo{year}{1992}\natexlab{}.
\newblock \showarticletitle{Geometric manipulation of tensor product surfaces}. In \bibinfo{booktitle}{\emph{Proceedings of the 1992 symposium on Interactive 3D graphics}}. \bibinfo{pages}{101--108}.
\newblock


\bibitem[Fu et~al\mbox{.}(2004)]%
        {fu2004topology}
\bibfield{author}{\bibinfo{person}{Hongbo Fu}, \bibinfo{person}{Chiew-Lan Tai}, {and} \bibinfo{person}{Hongxin Zhang}.} \bibinfo{year}{2004}\natexlab{}.
\newblock \showarticletitle{Topology-free cut-and-paste editing over meshes}. In \bibinfo{booktitle}{\emph{Geometric Modeling and Processing, 2004. Proceedings}}. IEEE, \bibinfo{pages}{173--182}.
\newblock


\bibitem[Galyean and Hughes(1991)]%
        {galyean1991sculpting}
\bibfield{author}{\bibinfo{person}{Tinsley~A Galyean} {and} \bibinfo{person}{John~F Hughes}.} \bibinfo{year}{1991}\natexlab{}.
\newblock \showarticletitle{Sculpting: An interactive volumetric modeling technique}.
\newblock \bibinfo{journal}{\emph{ACM SIGGRAPH Computer Graphics}} \bibinfo{volume}{25}, \bibinfo{number}{4} (\bibinfo{year}{1991}), \bibinfo{pages}{267--274}.
\newblock


\bibitem[Gourmel et~al\mbox{.}(2013)]%
        {gourmel2013gradient}
\bibfield{author}{\bibinfo{person}{Olivier Gourmel}, \bibinfo{person}{Loic Barthe}, \bibinfo{person}{Marie-Paule Cani}, \bibinfo{person}{Brian Wyvill}, \bibinfo{person}{Adrien Bernhardt}, \bibinfo{person}{Mathias Paulin}, {and} \bibinfo{person}{Herbert Grasberger}.} \bibinfo{year}{2013}\natexlab{}.
\newblock \showarticletitle{A gradient-based implicit blend}.
\newblock \bibinfo{journal}{\emph{ACM Transactions on Graphics (TOG)}} \bibinfo{volume}{32}, \bibinfo{number}{2} (\bibinfo{year}{2013}), \bibinfo{pages}{1--12}.
\newblock


\bibitem[Hall and Meyer(1976)]%
        {hall1976optimal}
\bibfield{author}{\bibinfo{person}{Charles~A Hall} {and} \bibinfo{person}{W~Weston Meyer}.} \bibinfo{year}{1976}\natexlab{}.
\newblock \showarticletitle{Optimal error bounds for cubic spline interpolation}.
\newblock \bibinfo{journal}{\emph{Journal of Approximation Theory}} \bibinfo{volume}{16}, \bibinfo{number}{2} (\bibinfo{year}{1976}), \bibinfo{pages}{105--122}.
\newblock


\bibitem[Hart(1996)]%
        {hart1996sphere}
\bibfield{author}{\bibinfo{person}{John~C Hart}.} \bibinfo{year}{1996}\natexlab{}.
\newblock \showarticletitle{Sphere tracing: A geometric method for the antialiased ray tracing of implicit surfaces}.
\newblock \bibinfo{journal}{\emph{The Visual Computer}} \bibinfo{volume}{12}, \bibinfo{number}{10} (\bibinfo{year}{1996}), \bibinfo{pages}{527--545}.
\newblock


\bibitem[Hart(1998)]%
        {hart1998morse}
\bibfield{author}{\bibinfo{person}{John~C Hart}.} \bibinfo{year}{1998}\natexlab{}.
\newblock \showarticletitle{Morse theory for implicit surface modeling}.
\newblock In \bibinfo{booktitle}{\emph{Mathematical Visualization}}. \bibinfo{publisher}{Springer}, \bibinfo{pages}{257--268}.
\newblock


\bibitem[Hestenes et~al\mbox{.}(1952)]%
        {hestenes1952methods}
\bibfield{author}{\bibinfo{person}{Magnus~R Hestenes}, \bibinfo{person}{Eduard Stiefel}, {et~al\mbox{.}}} \bibinfo{year}{1952}\natexlab{}.
\newblock \showarticletitle{Methods of conjugate gradients for solving linear systems}.
\newblock \bibinfo{journal}{\emph{Journal of research of the National Bureau of Standards}} \bibinfo{volume}{49}, \bibinfo{number}{6} (\bibinfo{year}{1952}), \bibinfo{pages}{409--436}.
\newblock


\bibitem[Hu and Wen(2019)]%
        {hu2019research}
\bibfield{author}{\bibinfo{person}{Ping Hu} {and} \bibinfo{person}{Jian Wen}.} \bibinfo{year}{2019}\natexlab{}.
\newblock \showarticletitle{Research on 3D animation character design based on multimedia interaction}.
\newblock \bibinfo{journal}{\emph{Multimedia Tools and Applications}} (\bibinfo{year}{2019}), \bibinfo{pages}{1--14}.
\newblock


\bibitem[Huang et~al\mbox{.}(2019)]%
        {huang2019variational}
\bibfield{author}{\bibinfo{person}{Zhiyang Huang}, \bibinfo{person}{Nathan Carr}, {and} \bibinfo{person}{Tao Ju}.} \bibinfo{year}{2019}\natexlab{}.
\newblock \showarticletitle{Variational implicit point set surfaces}.
\newblock \bibinfo{journal}{\emph{ACM Transactions on Graphics (TOG)}} \bibinfo{volume}{38}, \bibinfo{number}{4} (\bibinfo{year}{2019}), \bibinfo{pages}{1--13}.
\newblock


\bibitem[Ito et~al\mbox{.}(2021)]%
        {ito2021over}
\bibfield{author}{\bibinfo{person}{Tomohiko Ito}, \bibinfo{person}{Teruyoshi Kaneko}, \bibinfo{person}{Yoshiki Tanaka}, {and} \bibinfo{person}{Sato Saga}.} \bibinfo{year}{2021}\natexlab{}.
\newblock \showarticletitle{Over-sketching operation to realize geometrical and topological editing across multiple objects in sketch-based CAD interface}. In \bibinfo{booktitle}{\emph{26th International Conference on Intelligent User Interfaces-Companion}}. \bibinfo{pages}{49--51}.
\newblock


\bibitem[Jacobson et~al\mbox{.}(2013)]%
        {jacobson2013robust}
\bibfield{author}{\bibinfo{person}{Alec Jacobson}, \bibinfo{person}{Ladislav Kavan}, {and} \bibinfo{person}{Olga Sorkine-Hornung}.} \bibinfo{year}{2013}\natexlab{}.
\newblock \showarticletitle{Robust inside-outside segmentation using generalized winding numbers}.
\newblock \bibinfo{journal}{\emph{ACM Transactions on Graphics (TOG)}} \bibinfo{volume}{32}, \bibinfo{number}{4} (\bibinfo{year}{2013}), \bibinfo{pages}{1--12}.
\newblock


\bibitem[Jacobson et~al\mbox{.}(2018)]%
        {libigl}
\bibfield{author}{\bibinfo{person}{Alec Jacobson}, \bibinfo{person}{Daniele Panozzo}, {et~al\mbox{.}}} \bibinfo{year}{2018}\natexlab{}.
\newblock \bibinfo{title}{{libigl}: A simple {C++} geometry processing library}.
\newblock
\newblock
\newblock
\shownote{https://libigl.github.io/}.


\bibitem[Ju et~al\mbox{.}(2007)]%
        {ju2007editing}
\bibfield{author}{\bibinfo{person}{Tao Ju}, \bibinfo{person}{Qian-Yi Zhou}, {and} \bibinfo{person}{Shi-Min Hu}.} \bibinfo{year}{2007}\natexlab{}.
\newblock \showarticletitle{Editing the topology of 3D models by sketching}.
\newblock \bibinfo{journal}{\emph{ACM Transactions on Graphics (TOG)}} \bibinfo{volume}{26}, \bibinfo{number}{3} (\bibinfo{year}{2007}), \bibinfo{pages}{42--es}.
\newblock


\bibitem[Kazhdan and Hoppe(2013)]%
        {kazhdan2013screened}
\bibfield{author}{\bibinfo{person}{Michael Kazhdan} {and} \bibinfo{person}{Hugues Hoppe}.} \bibinfo{year}{2013}\natexlab{}.
\newblock \showarticletitle{Screened poisson surface reconstruction}.
\newblock \bibinfo{journal}{\emph{ACM Transactions on Graphics (ToG)}} \bibinfo{volume}{32}, \bibinfo{number}{3} (\bibinfo{year}{2013}), \bibinfo{pages}{1--13}.
\newblock


\bibitem[Koch et~al\mbox{.}(2019)]%
        {ABC}
\bibfield{author}{\bibinfo{person}{Sebastian Koch}, \bibinfo{person}{Albert Matveev}, \bibinfo{person}{Zhongshi Jiang}, \bibinfo{person}{Francis Williams}, \bibinfo{person}{Alexey Artemov}, \bibinfo{person}{Evgeny Burnaev}, \bibinfo{person}{Marc Alexa}, \bibinfo{person}{Denis Zorin}, {and} \bibinfo{person}{Daniele Panozzo}.} \bibinfo{year}{2019}\natexlab{}.
\newblock \showarticletitle{ABC: A Big CAD Model Dataset For Geometric Deep Learning}. In \bibinfo{booktitle}{\emph{The IEEE Conference on Computer Vision and Pattern Recognition (CVPR)}}.
\newblock


\bibitem[lin Yu and Tsao(2022)]%
        {lin2022study}
\bibfield{author}{\bibinfo{person}{Kai lin Yu} {and} \bibinfo{person}{Yung-Chin Tsao}.} \bibinfo{year}{2022}\natexlab{}.
\newblock \showarticletitle{A STUDY OF CHARACTER DESIGN METHOD.}
\newblock \bibinfo{journal}{\emph{International Journal of Organizational Innovation}} \bibinfo{volume}{15}, \bibinfo{number}{1} (\bibinfo{year}{2022}).
\newblock


\bibitem[Mandal et~al\mbox{.}(2022)]%
        {mandal2022interactive}
\bibfield{author}{\bibinfo{person}{Avirup Mandal}, \bibinfo{person}{Parag Chaudhuri}, {and} \bibinfo{person}{Subhasis Chaudhuri}.} \bibinfo{year}{2022}\natexlab{}.
\newblock \showarticletitle{Interactive physics-based virtual sculpting with haptic feedback}.
\newblock \bibinfo{journal}{\emph{Proceedings of the ACM on Computer Graphics and Interactive Techniques}} \bibinfo{volume}{5}, \bibinfo{number}{1} (\bibinfo{year}{2022}), \bibinfo{pages}{1--20}.
\newblock


\bibitem[Moo-Young et~al\mbox{.}(2021)]%
        {moo2021virtual}
\bibfield{author}{\bibinfo{person}{Joss~Kingdom Moo-Young}, \bibinfo{person}{Andrew Hogue}, {and} \bibinfo{person}{Veronika Szkudlarek}.} \bibinfo{year}{2021}\natexlab{}.
\newblock \showarticletitle{Virtual materiality: Realistic clay sculpting in vr}. In \bibinfo{booktitle}{\emph{Extended Abstracts of the 2021 Annual Symposium on Computer-Human Interaction in Play}}. \bibinfo{pages}{105--110}.
\newblock


\bibitem[Mourrain and Pavone(2009)]%
        {mourrain2009subdivision}
\bibfield{author}{\bibinfo{person}{Bernard Mourrain} {and} \bibinfo{person}{Jean~Pascal Pavone}.} \bibinfo{year}{2009}\natexlab{}.
\newblock \showarticletitle{Subdivision methods for solving polynomial equations}.
\newblock \bibinfo{journal}{\emph{Journal of Symbolic Computation}} \bibinfo{volume}{44}, \bibinfo{number}{3} (\bibinfo{year}{2009}), \bibinfo{pages}{292--306}.
\newblock


\bibitem[Museth et~al\mbox{.}(2002)]%
        {museth2002level}
\bibfield{author}{\bibinfo{person}{Ken Museth}, \bibinfo{person}{David~E Breen}, \bibinfo{person}{Ross~T Whitaker}, {and} \bibinfo{person}{Alan~H Barr}.} \bibinfo{year}{2002}\natexlab{}.
\newblock \showarticletitle{Level set surface editing operators}. In \bibinfo{booktitle}{\emph{Proceedings of the 29th annual conference on Computer graphics and interactive techniques}}. \bibinfo{pages}{330--338}.
\newblock


\bibitem[Museth et~al\mbox{.}(2005)]%
        {museth2005algorithms}
\bibfield{author}{\bibinfo{person}{Ken Museth}, \bibinfo{person}{David~E Breen}, \bibinfo{person}{Ross~T Whitaker}, \bibinfo{person}{Sean Mauch}, {and} \bibinfo{person}{David Johnson}.} \bibinfo{year}{2005}\natexlab{}.
\newblock \showarticletitle{Algorithms for interactive editing of level set models}. In \bibinfo{booktitle}{\emph{Computer Graphics Forum}}, Vol.~\bibinfo{volume}{24}. Wiley Online Library, \bibinfo{pages}{821--841}.
\newblock


\bibitem[Pasko et~al\mbox{.}(2005)]%
        {pasko2005bounded}
\bibfield{author}{\bibinfo{person}{Galina~I Pasko}, \bibinfo{person}{Alexander~A Pasko}, {and} \bibinfo{person}{Tosiyasu~L Kunii}.} \bibinfo{year}{2005}\natexlab{}.
\newblock \showarticletitle{Bounded blending for function-based shape modeling}.
\newblock \bibinfo{journal}{\emph{IEEE Computer Graphics and Applications}} \bibinfo{volume}{25}, \bibinfo{number}{2} (\bibinfo{year}{2005}), \bibinfo{pages}{36--45}.
\newblock


\bibitem[Peng et~al\mbox{.}(2018)]%
        {peng2018autocomplete}
\bibfield{author}{\bibinfo{person}{Mengqi Peng}, \bibinfo{person}{Jun Xing}, {and} \bibinfo{person}{Li-Yi Wei}.} \bibinfo{year}{2018}\natexlab{}.
\newblock \showarticletitle{Autocomplete 3d sculpting}.
\newblock \bibinfo{journal}{\emph{ACM Transactions on Graphics (TOG)}} \bibinfo{volume}{37}, \bibinfo{number}{4} (\bibinfo{year}{2018}), \bibinfo{pages}{1--15}.
\newblock


\bibitem[Perry and Frisken(2001)]%
        {perry2001kizamu}
\bibfield{author}{\bibinfo{person}{Ronald~N Perry} {and} \bibinfo{person}{Sarah~F Frisken}.} \bibinfo{year}{2001}\natexlab{}.
\newblock \showarticletitle{Kizamu: A system for sculpting digital characters}. In \bibinfo{booktitle}{\emph{Proceedings of the 28th annual conference on Computer graphics and interactive techniques}}. \bibinfo{pages}{47--56}.
\newblock


\bibitem[Rouhani et~al\mbox{.}(2014)]%
        {rouhani2014implicit}
\bibfield{author}{\bibinfo{person}{Mohammad Rouhani}, \bibinfo{person}{Angel~D Sappa}, {and} \bibinfo{person}{Edmond Boyer}.} \bibinfo{year}{2014}\natexlab{}.
\newblock \showarticletitle{Implicit B-spline surface reconstruction}.
\newblock \bibinfo{journal}{\emph{IEEE Transactions on Image Processing}} \bibinfo{volume}{24}, \bibinfo{number}{1} (\bibinfo{year}{2014}), \bibinfo{pages}{22--32}.
\newblock


\bibitem[Schmidt et~al\mbox{.}(2005)]%
        {schmidt2005interactive}
\bibfield{author}{\bibinfo{person}{Ryan Schmidt}, \bibinfo{person}{Brian Wyvill}, {and} \bibinfo{person}{Eric Galin}.} \bibinfo{year}{2005}\natexlab{}.
\newblock \showarticletitle{Interactive implicit modeling with hierarchical spatial caching}. In \bibinfo{booktitle}{\emph{International Conference on Shape Modeling and Applications 2005 (SMI'05)}}. IEEE, \bibinfo{pages}{104--113}.
\newblock


\bibitem[Sharf et~al\mbox{.}(2007)]%
        {sharf2007interactive}
\bibfield{author}{\bibinfo{person}{Andrei Sharf}, \bibinfo{person}{Thomas Lewiner}, \bibinfo{person}{Gil Shklarski}, \bibinfo{person}{Sivan Toledo}, {and} \bibinfo{person}{Daniel Cohen-Or}.} \bibinfo{year}{2007}\natexlab{}.
\newblock \showarticletitle{Interactive topology-aware surface reconstruction}.
\newblock \bibinfo{journal}{\emph{ACM Transactions on Graphics (TOG)}} \bibinfo{volume}{26}, \bibinfo{number}{3} (\bibinfo{year}{2007}), \bibinfo{pages}{43--es}.
\newblock


\bibitem[Stander and Hart(1997)]%
        {stander1997guaranteeing}
\bibfield{author}{\bibinfo{person}{Barton~T. Stander} {and} \bibinfo{person}{John~C. Hart}.} \bibinfo{year}{1997}\natexlab{}.
\newblock \showarticletitle{Guaranteeing the Topology of an Implicit Surface Polygonization for Interactive Modeling}. In \bibinfo{booktitle}{\emph{Proceedings of the 24th Annual Conference on Computer Graphics and Interactive Techniques}} \emph{(\bibinfo{series}{SIGGRAPH '97})}. \bibinfo{publisher}{ACM Press/Addison-Wesley Publishing Co.}, \bibinfo{address}{USA}, \bibinfo{pages}{279–286}.
\newblock
\showISBNx{0897918967}
\urldef\tempurl%
\url{https://doi.org/10.1145/258734.258868}
\showDOI{\tempurl}


\bibitem[Tang and Feng(2018)]%
        {tang2018multi}
\bibfield{author}{\bibinfo{person}{Yizhi Tang} {and} \bibinfo{person}{Jieqing Feng}.} \bibinfo{year}{2018}\natexlab{}.
\newblock \showarticletitle{Multi-scale surface reconstruction based on a curvature-adaptive signed distance field}.
\newblock \bibinfo{journal}{\emph{Computers \& Graphics}}  \bibinfo{volume}{70} (\bibinfo{year}{2018}), \bibinfo{pages}{28--38}.
\newblock


\bibitem[Terzopoulos and Qin(1994)]%
        {terzopoulos1994dynamic}
\bibfield{author}{\bibinfo{person}{Demetri Terzopoulos} {and} \bibinfo{person}{Hong Qin}.} \bibinfo{year}{1994}\natexlab{}.
\newblock \showarticletitle{Dynamic NURBS with geometric constraints for interactive sculpting}.
\newblock \bibinfo{journal}{\emph{ACM Transactions on Graphics (TOG)}} \bibinfo{volume}{13}, \bibinfo{number}{2} (\bibinfo{year}{1994}), \bibinfo{pages}{103--136}.
\newblock


\bibitem[Wong et~al\mbox{.}(2000)]%
        {wong2000virtual}
\bibfield{author}{\bibinfo{person}{Janis~PY Wong}, \bibinfo{person}{Rynson~WH Lau}, {and} \bibinfo{person}{Lizhuang Ma}.} \bibinfo{year}{2000}\natexlab{}.
\newblock \showarticletitle{Virtual 3d sculpting}.
\newblock \bibinfo{journal}{\emph{The Journal of Visualization and Computer Animation}} \bibinfo{volume}{11}, \bibinfo{number}{3} (\bibinfo{year}{2000}), \bibinfo{pages}{155--166}.
\newblock


\bibitem[Xiong and Zimin(2022)]%
        {xiong2022algorithm}
\bibfield{author}{\bibinfo{person}{Huang Xiong} {and} \bibinfo{person}{Yao Zimin}.} \bibinfo{year}{2022}\natexlab{}.
\newblock \showarticletitle{Algorithm modeling technology of computer aided fractal art pattern design}. In \bibinfo{booktitle}{\emph{2022 World Automation Congress (WAC)}}. IEEE, \bibinfo{pages}{62--65}.
\newblock


\bibitem[Yin et~al\mbox{.}(2014)]%
        {yin2014morfit}
\bibfield{author}{\bibinfo{person}{Kangxue Yin}, \bibinfo{person}{Hui Huang}, \bibinfo{person}{Hao Zhang}, \bibinfo{person}{Minglun Gong}, \bibinfo{person}{Daniel Cohen-Or}, {and} \bibinfo{person}{Baoquan Chen}.} \bibinfo{year}{2014}\natexlab{}.
\newblock \showarticletitle{Morfit: interactive surface reconstruction from incomplete point clouds with curve-driven topology and geometry control.}
\newblock \bibinfo{journal}{\emph{ACM Trans. Graph.}} \bibinfo{volume}{33}, \bibinfo{number}{6} (\bibinfo{year}{2014}), \bibinfo{pages}{202--1}.
\newblock


\bibitem[Zhou and Jacobson(2016)]%
        {Thingi10K}
\bibfield{author}{\bibinfo{person}{Qingnan Zhou} {and} \bibinfo{person}{Alec Jacobson}.} \bibinfo{year}{2016}\natexlab{}.
\newblock \showarticletitle{Thingi10K: A Dataset of 10,000 3D-Printing Models}.
\newblock \bibinfo{journal}{\emph{arXiv preprint arXiv:1605.04797}} (\bibinfo{year}{2016}).
\newblock


\end{thebibliography}

\clearpage

\begin{table*}[!t]
\centering
\caption{Quantitative comparison for topology editing.
We present the average values to assess the results produced by volunteers. 
}
\Description{The differences between the results of four tasks completed using our method and STEM, with respect to various metrics, in comparison to the target.}
\label{tab:quan_topology}

\resizebox{.9\linewidth}{!}{%
\begin{tabular}{l|cc|cc|cc|ccc|ccc}
\toprule
       & \multicolumn{2}{c|}{Chamfer~$\downarrow$}              & \multicolumn{2}{c|}{F-Score~$\uparrow$}          & \multicolumn{2}{c|}{Normal C.~$\uparrow$}        & \multicolumn{3}{c|}{Genus}     & \multicolumn{3}{c}{Components} \\
       & Ours          & STEM~\shortcite{ju2007editing}               & Ours        & STEM~\shortcite{ju2007editing}             & Ours        & STEM~\shortcite{ju2007editing}             & Ours  & STEM~\shortcite{ju2007editing}       & GT & Ours  & STEM~\shortcite{ju2007editing}       & GT \\ 
       \midrule
Task 1 & \textbf{2.80}         & 3.84         & \textbf{94.00}       & 91.46       & \textbf{98.68}       & 94.50       & \textbf{3.14} & 3.20         & 3  & \textbf{1.00}           & 1.43 & 1  \\
Task 2 & \textbf{2.86} & 3.72 & \textbf{93.62}  & 84.21 & \textbf{98.02} & 94.14 & \textbf{8.15} & 7.67 & 8  & \textbf{1.08} & 1.77 & 1  \\
Task 3 & \textbf{2.96} & 4.33 & \textbf{92.78} & 84.41 & \textbf{96.69} & 92.26 & \textbf{1.00}           & 5.54    & 1  & \textbf{1.00}           & 1.13    & 1  \\
Task 4 & \textbf{2.65} & 4.14 & \textbf{94.36} & 79.40 & \textbf{99.22} & 94.58 & \textbf{1.00}           & 3.92 & 1  & \textbf{1.00}           & \textbf{1.00}           & 1  \\ 
\bottomrule
\end{tabular}
}
\end{table*}

\begin{figure}[h]
  \centering
  \begin{subfigure}{\linewidth}
    \includegraphics[width=\linewidth]{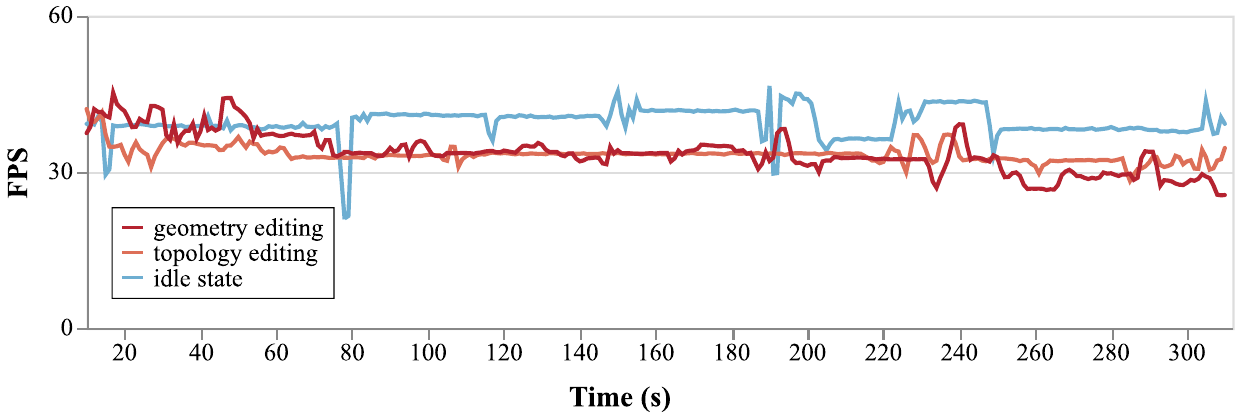}
    \vspace{-5mm}
    \caption{} 
    \Description{The FPS values of approximately 30 on the GTX 1060.}
    \label{fig:performance_fpsa}
  \end{subfigure}

  \begin{subfigure}{\linewidth}
    \includegraphics[width=\linewidth]{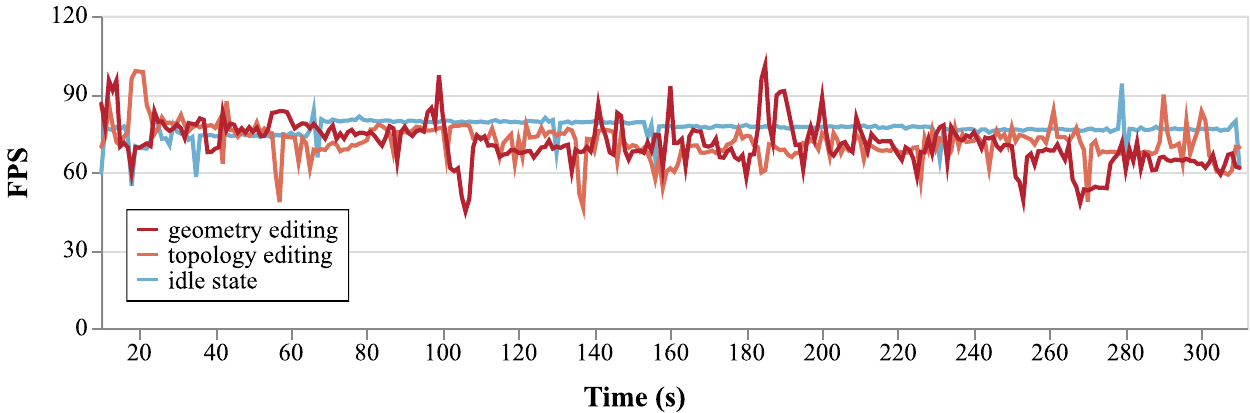}
    \vspace{-5mm}
    \caption{} 
    \Description{The FPS values of approximately 60 on the RTX 3060.}
    \label{fig:performance_fpsb}
  \end{subfigure}
 \vspace{-6mm}
 \caption{FPS statistical results for topology editing, geometry editing, and idle state on NVIDIA GeForce GTX 1060~(a) and NVIDIA GeForce RTX 3060~(b) GPUs.}
 \Description{The FPS values of approximately 30 on the GTX 1060 and around 60 on the RTX 3060.}
 \vspace{-5mm}
 \label{fig:performance_fps}
\end{figure}

\begin{figure}[H]
    \centering
    \begin{minipage}[t]{\linewidth}
    \centering
        \begin{minipage}[t]{0.31\linewidth}
        \centering
        \includegraphics[width=0.95\linewidth]{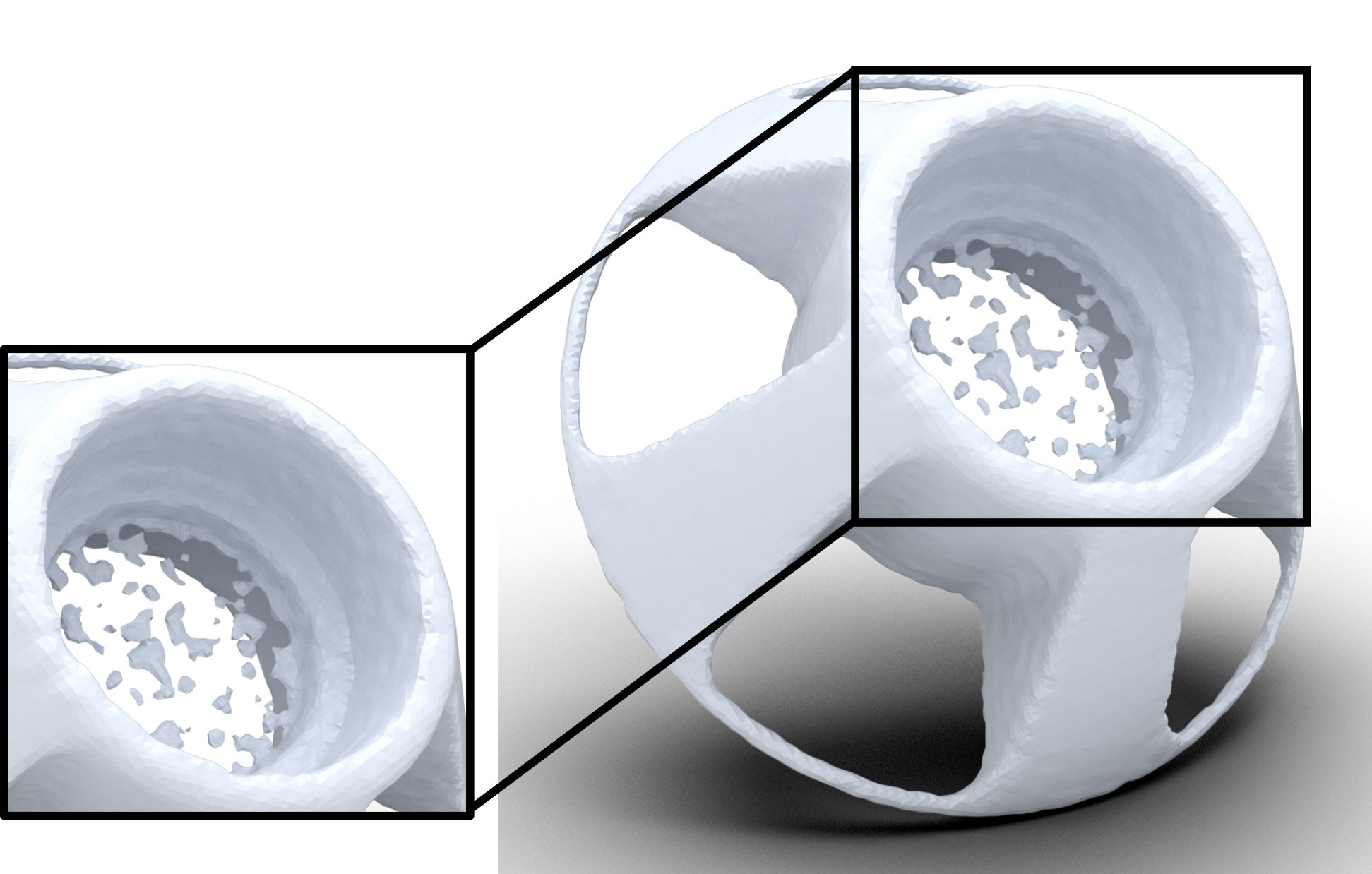}
        \Description{}
        \end{minipage}
        \begin{minipage}[t]{0.31\linewidth}
        \centering
        \includegraphics[width=0.95\linewidth]{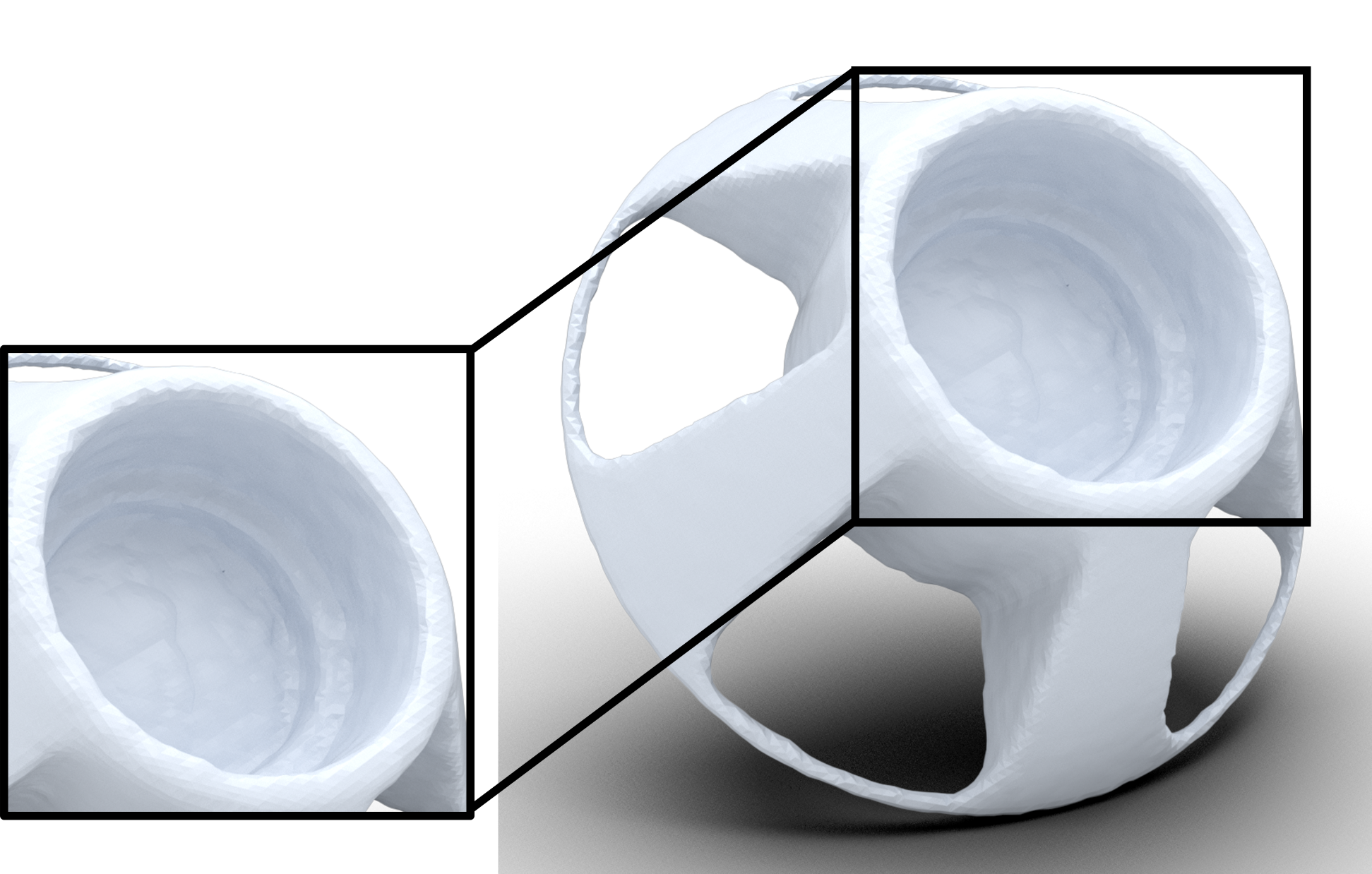}
        \Description{}
        \end{minipage}
        \begin{minipage}[t]{0.31\linewidth}
        \centering
        \includegraphics[width=0.95\linewidth]{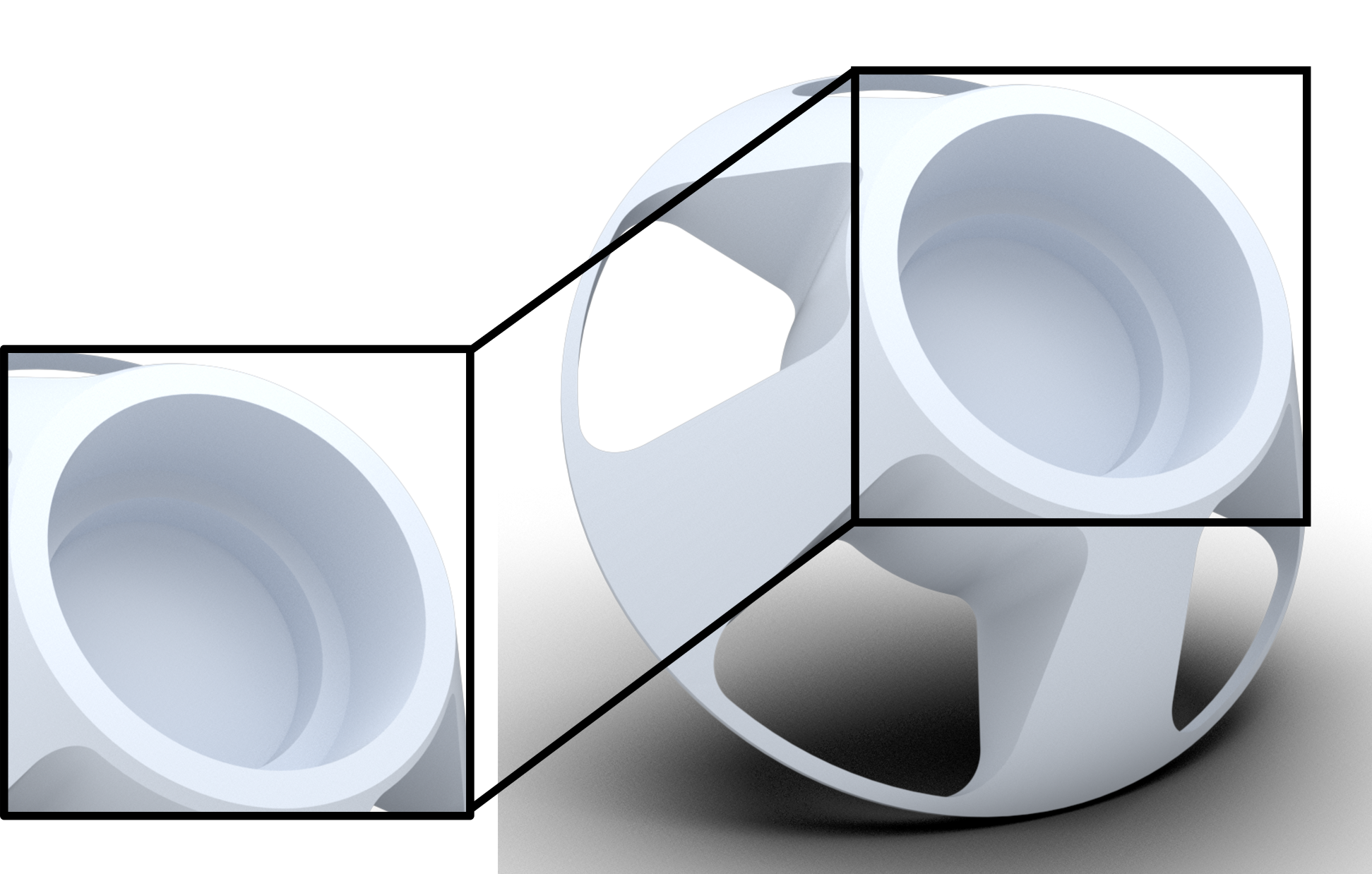}
        \Description{}
        \end{minipage} 
    \vspace{2mm}
    \end{minipage} 
        \begin{minipage}[t]{\linewidth}
    \centering
        \begin{minipage}[t]{0.31\linewidth}
        \centering
        \includegraphics[width=0.95\linewidth]{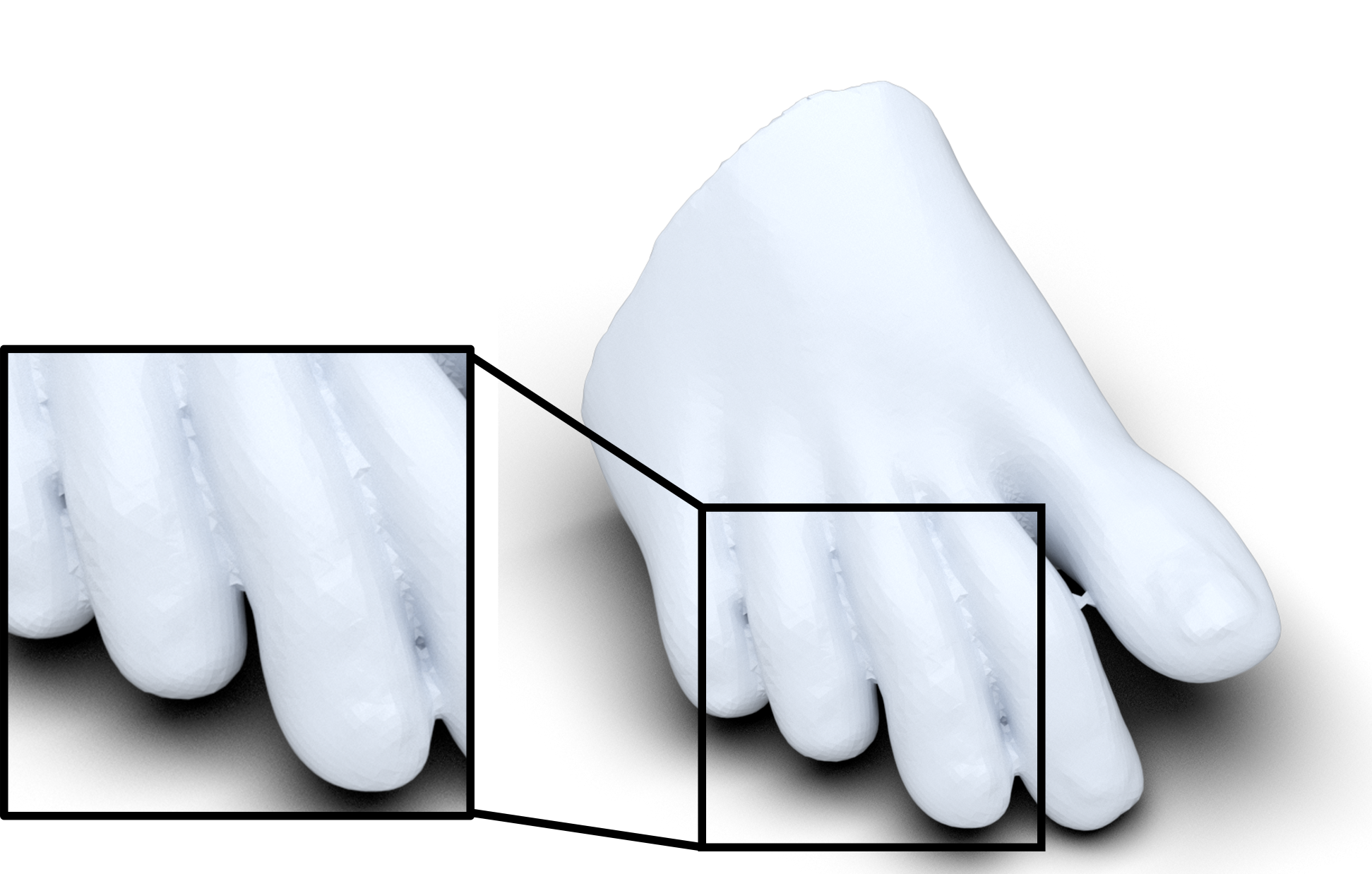}
        \Description{}
        \end{minipage}
        \begin{minipage}[t]{0.31\linewidth}
        \centering
        \includegraphics[width=0.95\linewidth]{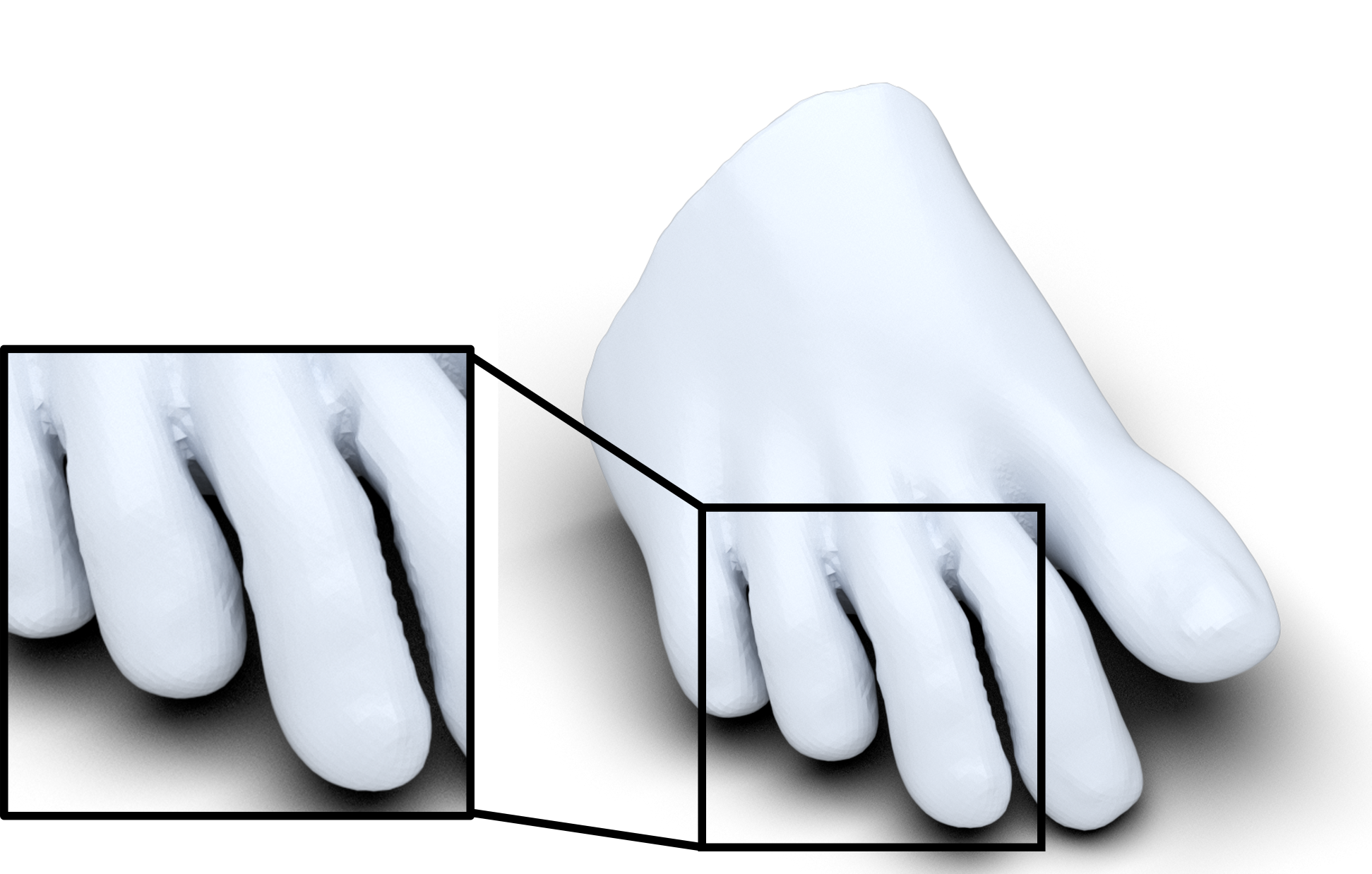}
        \Description{}
        \end{minipage}
        \begin{minipage}[t]{0.31\linewidth}
        \centering
        \includegraphics[width=0.95\linewidth]{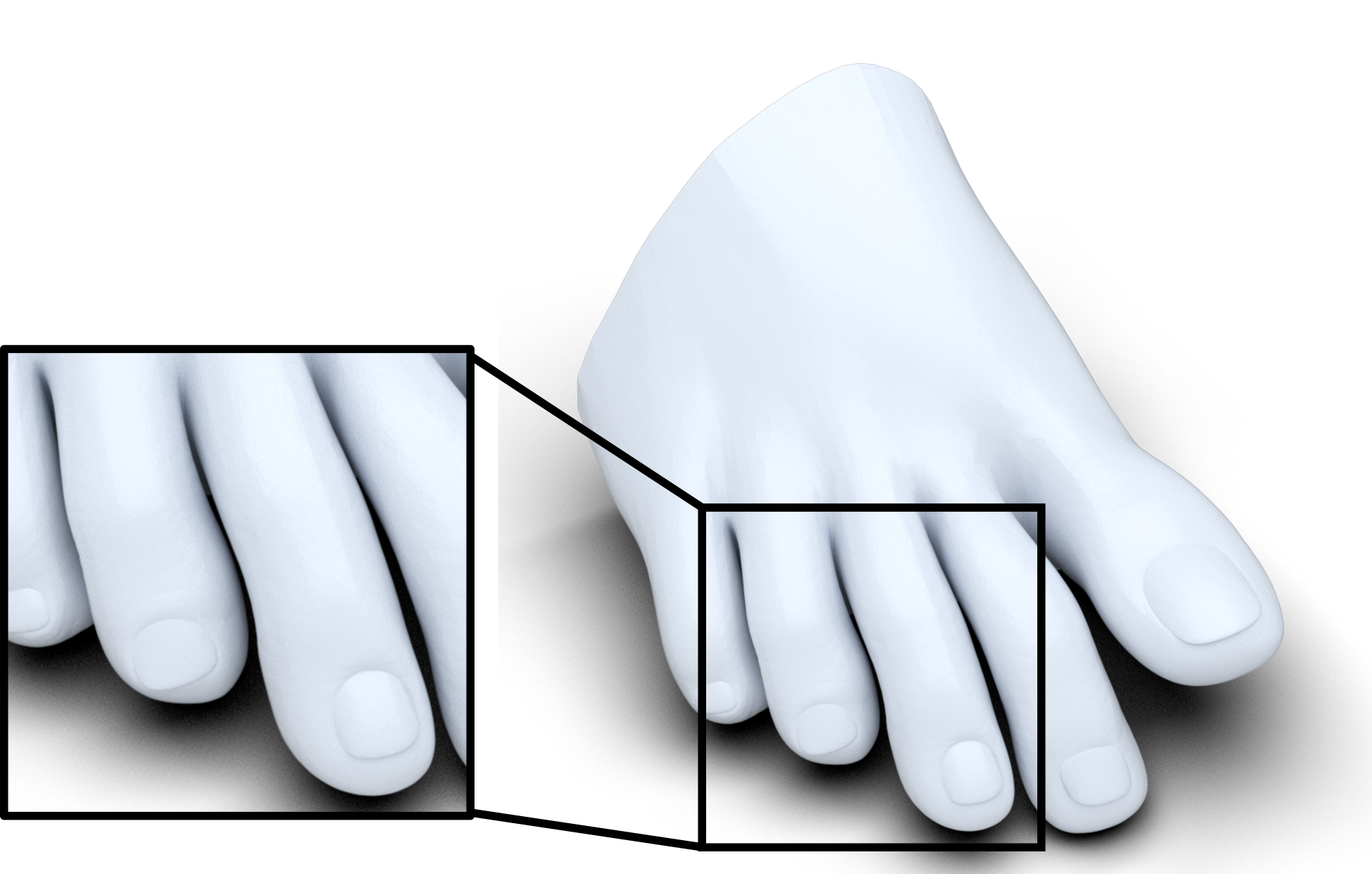}
        \Description{}
        \end{minipage} 
        \vspace{2mm}
    \end{minipage} 
   \begin{minipage}[t]{\linewidth}
    \centering
        \begin{minipage}[t]{0.31\linewidth}
        \centering
        \includegraphics[width=0.95\linewidth]{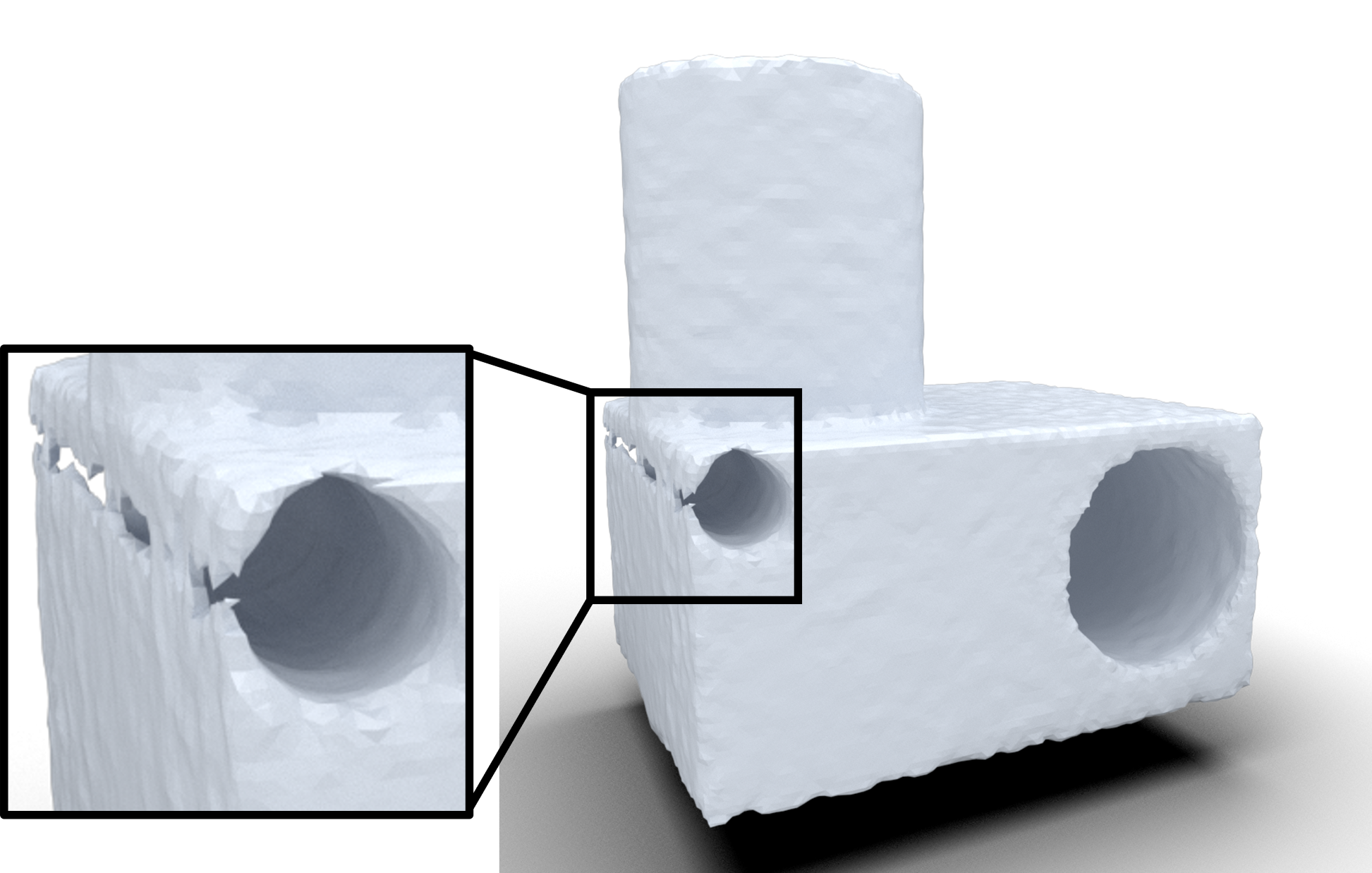}
        \Description{}
        \end{minipage}
        \begin{minipage}[t]{0.31\linewidth}
        \centering
        \includegraphics[width=0.95\linewidth]{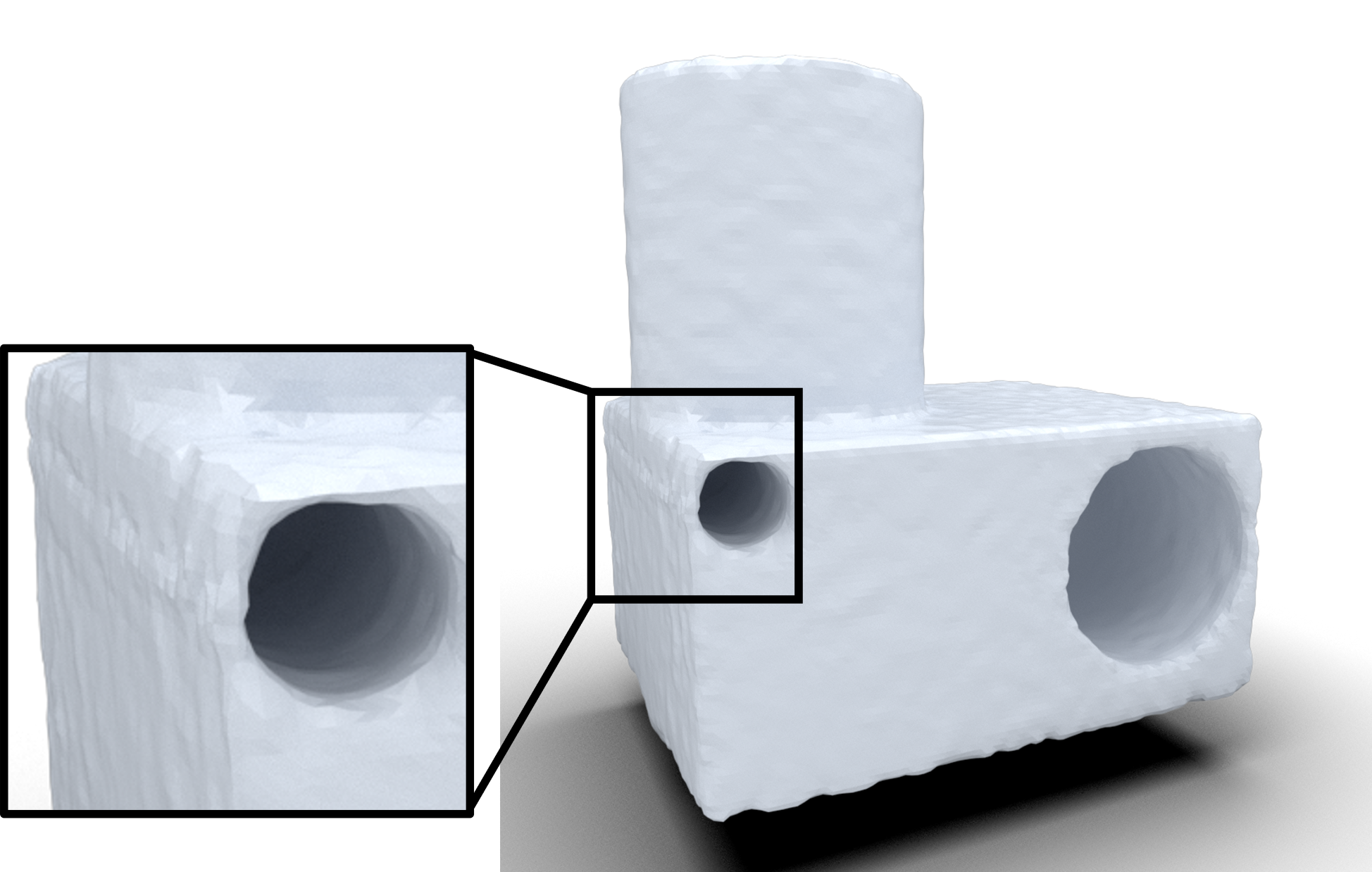}
        \Description{}
        \end{minipage}
        \begin{minipage}[t]{0.31\linewidth}
        \centering
        \includegraphics[width=0.95\linewidth]{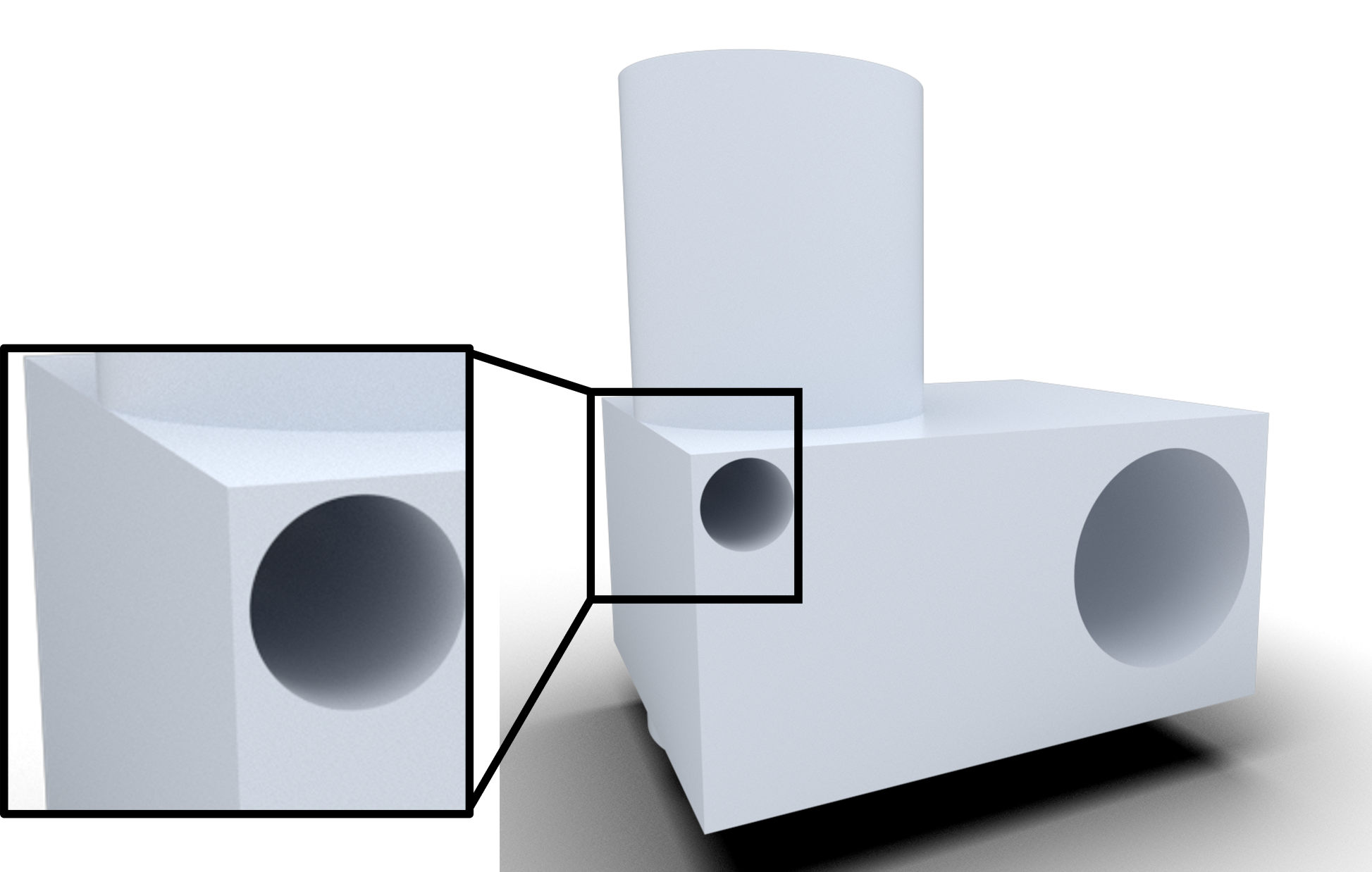}
        \Description{}
        \end{minipage}   
        \vspace{2mm}
    \end{minipage} 
    \begin{minipage}[t]{\linewidth}
    \centering
        \begin{minipage}[t]{0.31\linewidth}
        \captionsetup{labelformat=empty}
        \centering
        \includegraphics[width=0.95\linewidth]{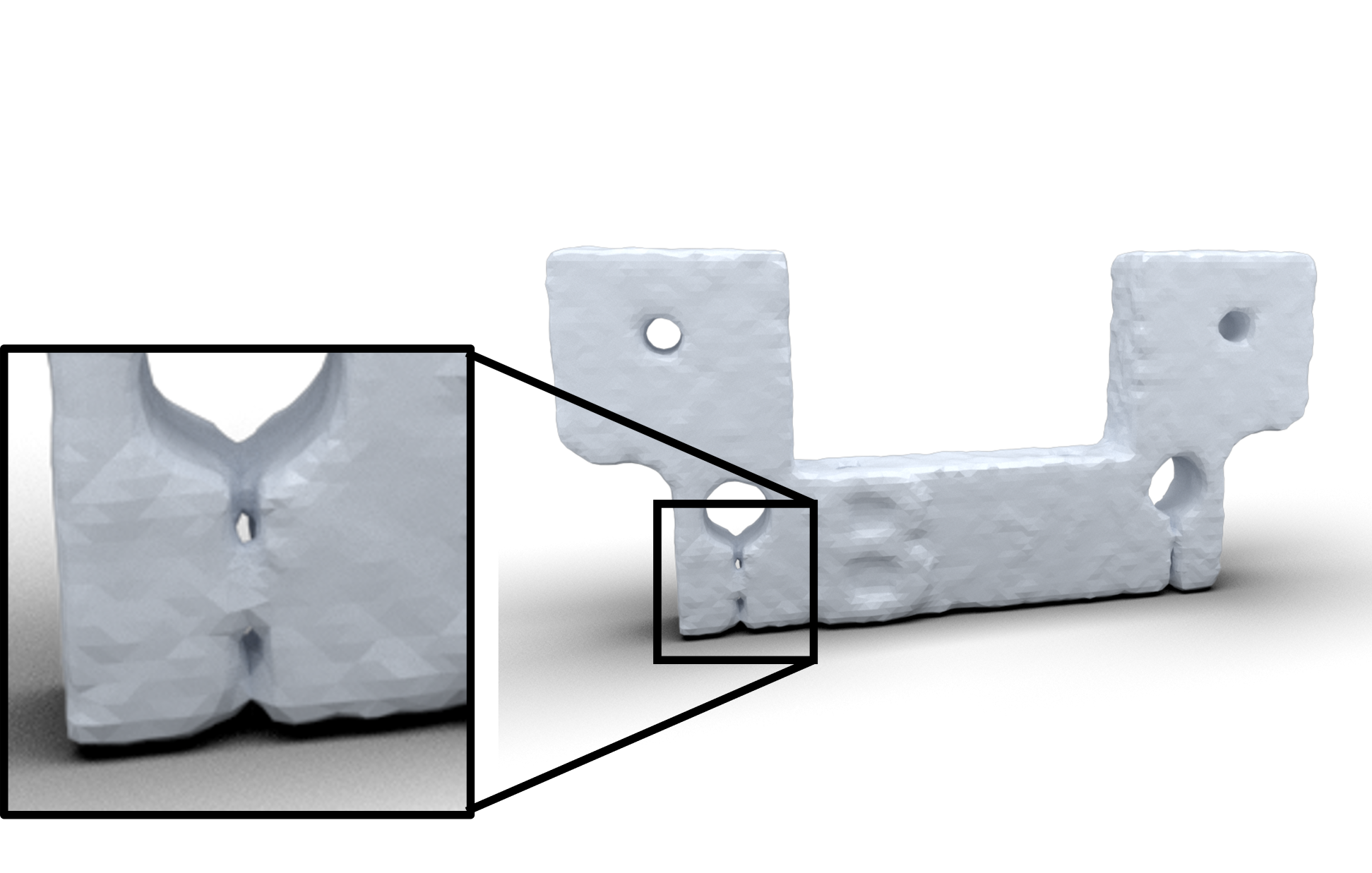}
        \caption{SPR}
        \Description{}
        \end{minipage}
        \begin{minipage}[t]{0.31\linewidth}
        \captionsetup{labelformat=empty}
        \centering
        \includegraphics[width=0.95\linewidth]{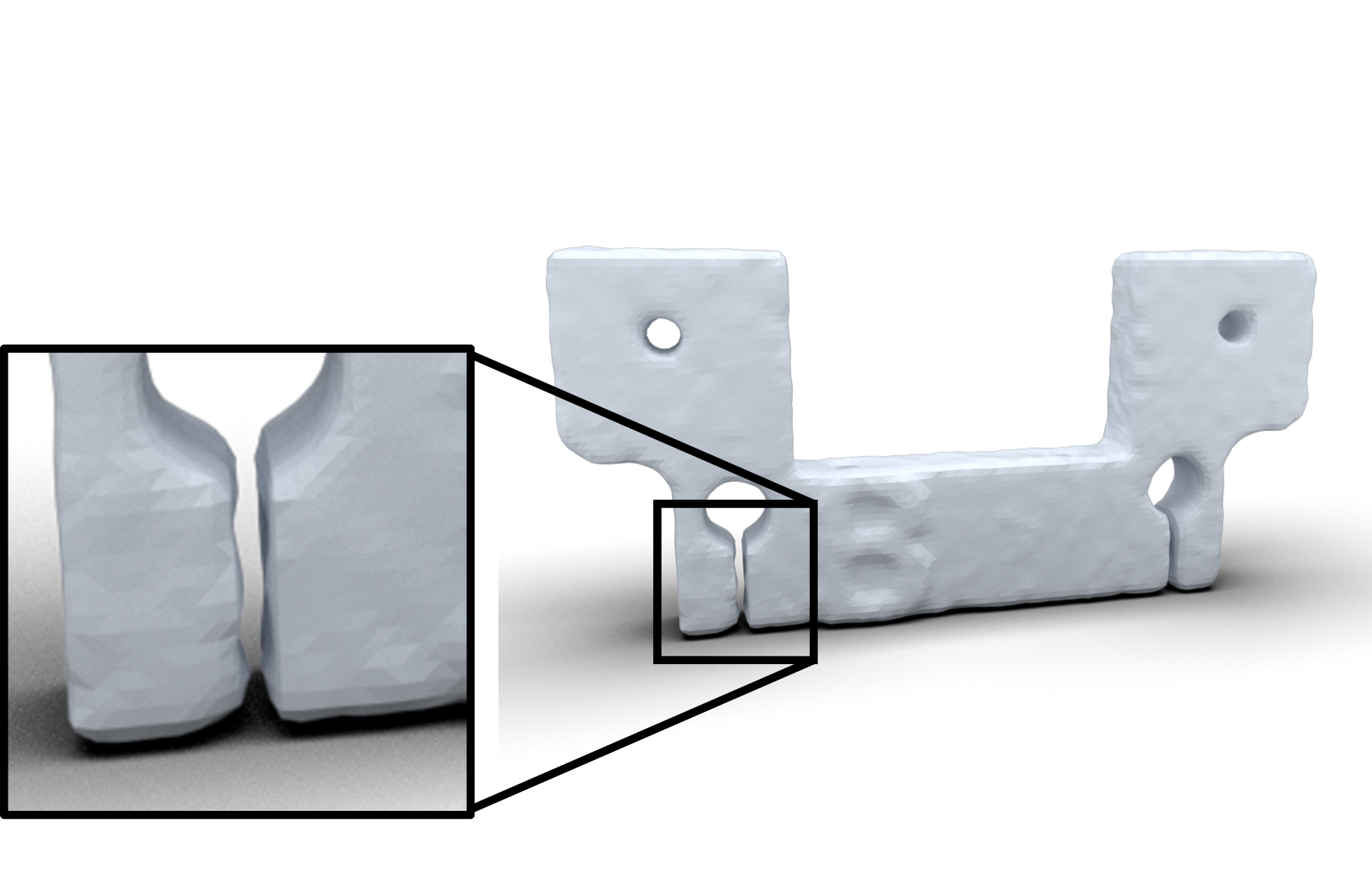}
        \caption{SPR $+$ Ours}
        \Description{}
        \end{minipage}
        \begin{minipage}[t]{0.31\linewidth}
        \captionsetup{labelformat=empty}
        \centering
        \includegraphics[width=0.95\linewidth]{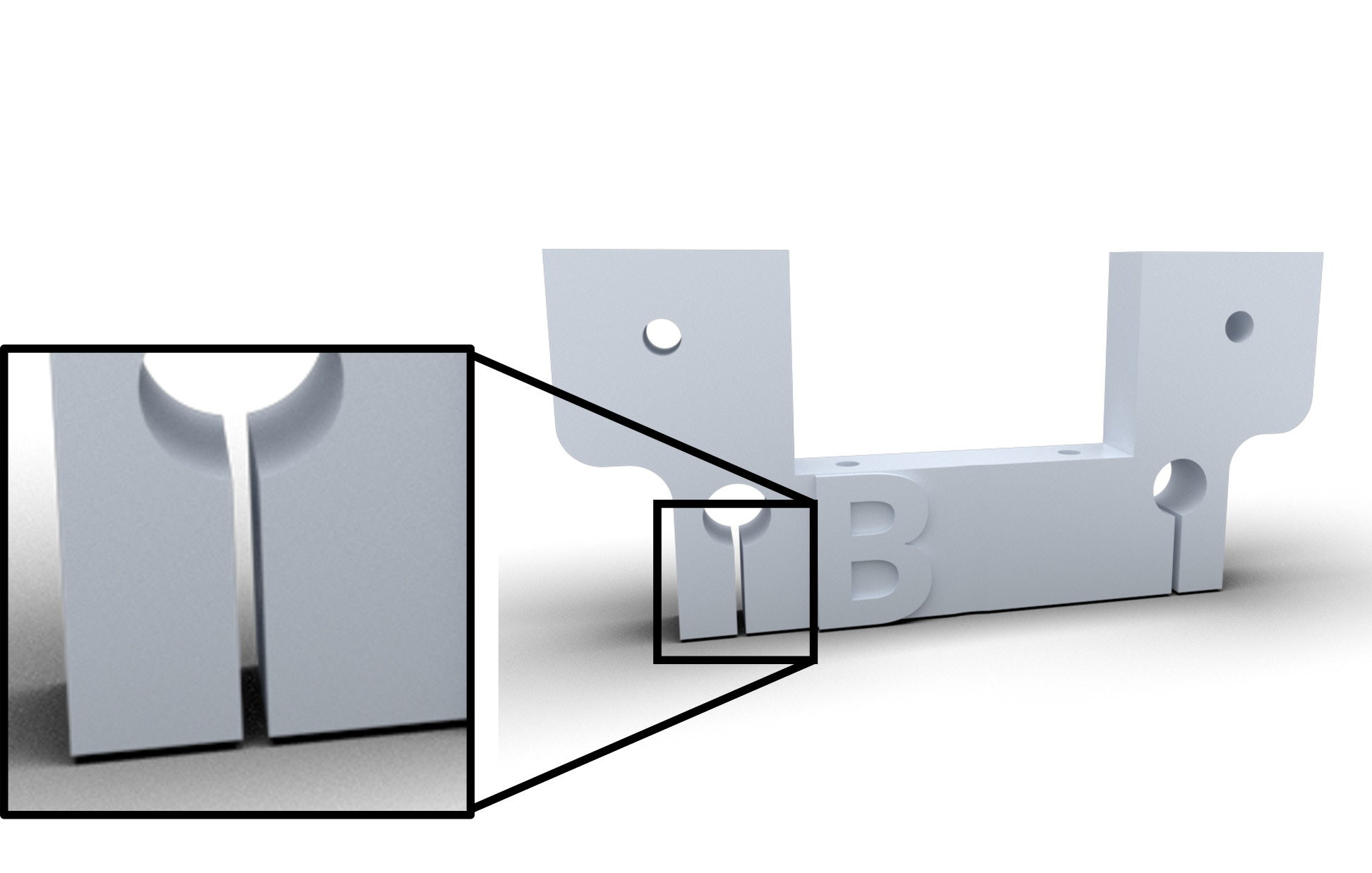}
        \caption{GT}
        \Description{}
        \end{minipage}
    \end{minipage}
    \setcounter{figure}{14}
    \caption{     
    We sampled 10K points from the ground-truth models (right) using random sampling, and reconstructed the surface using Screened Poisson Reconstruction~(SPR)~\cite{kazhdan2013screened} with ground truth normals. The reconstruction results (left) deviate from the ground truth shape and contain various structural artifacts due to point insufficiency. However, with our interactive system, these structural artifacts can be easily addressed (middle). It is worth noting that the four models are from the ABC~\cite{ABC} dataset and Thing10K~\cite{Thingi10K} dataset.
    }
    \vspace{-3mm}
    \Description{The topological editing results for the four models from the ABC dataset and Thing10K dataset.}
    \label{fig:app1}
\end{figure}

\end{document}